\documentclass[aps,pre,showpacs%
        ,amsmath%
        ,reprint%
        ]{revtex4-1}
\usepackage{graphicx}

\usepackage{amssymb}

\usepackage[mathscr]{eucal}

\newcommand{\set}[1]{\left\{ #1 \right\}}

\newcommand{\bra}[1]{\langle{}#1{}|}
\newcommand{\ket}[1]{|{}#1{}\rangle}
\newcommand{\bracket}[2]{\langle{}#1{}|{}#2{}\rangle}

\newcommand{\ketbra}[2]{|{}#1{}\rangle\langle{}#2{}|}
\newcommand{\bvec}[1]{\boldsymbol{#1}} 
\newcommand{\rmRe}{\mathop{\rm Re}}
\newcommand{\rmIm}{\mathop{\rm Im}}

\begin{document}


\title{Exotic quantum holonomy and higher-order exceptional points 
  \\
  in quantum kicked tops}

\author{Atushi Tanaka}
\homepage[]{\tt http://researchmap.jp/tanaka-atushi/}
\affiliation{Department of Physics, Tokyo Metropolitan University,
   Hachioji, Tokyo 192-0397, Japan}

\author{Sang Wook Kim}
\affiliation{Department of Physics Education, Pusan National University,
   Busan 609-735, South Korea}

\author{Taksu Cheon}
\homepage[]{\tt http://researchmap.jp/T_Zen/}
\affiliation{Laboratory of Physics, Kochi University of Technology,
  Tosa Yamada, Kochi 782-8502, Japan}


\begin{abstract}
The correspondence between exotic quantum holonomy that occurs in families of Hermitian cycles, and exceptional points (EPs) for non-Hermitian quantum theory is examined in quantum kicked tops. Under a suitable condition, an explicit expressions of the adiabatic parameter dependencies of quasienergies and stationary states, which exhibit anholonomies, are obtained. It is also shown that the quantum kicked tops with the complexified adiabatic parameter have a higher order EP, which is broken into lower order EPs with the application of small perturbations.  The stability of exotic holonomy against such bifurcation is demonstrated.
\end{abstract}

\pacs{05.45.Mt, 03.65.Vf, 03.67.-a}

\maketitle

\section{Introduction}
\label{sec:Introduction}

Along adiabatic time evolution, the state of a closed quantum system
stays
within an eigenspace of the Hamiltonian, 
when the system is initially prepared to be in
a stationary state~\cite{Kato-JPSJ-5-435}. It is nevertheless
possible that an adiabatic cycle induces nontrivial change. 
The most famous example 
reported by
Berry 
is
the appearance of the geometric phase
factor~\cite{Berry-PRSLA-392-45}. A non-Abelian version of the
quantum phase holonomy was subsequently reported by Wilczek and
Zee~\cite{Wilczek-PRL-52-2111}. Later, it was shown that an adiabatic
cycle can induce the interchanges of eigenenergies and
eigenspaces~\cite{Cheon-PLA-248-285}. 
Namely,
under the presence of such an exotic quantum holonomy, 
the initial and
final states of an adiabatic cycle belong to different eigenspaces.
The exotic quantum holonomy has been studied both in 
one-body~\cite{Tanaka-PRL-98-160407,Miyamoto-PRA-76-042115,Cheon-EPL-85-20001,Cheon-PLA-374-144,Tanaka-PRA-82-022104,Tanaka-EPL-96-10005}
and in many-body systems~\cite{Yonezawa-PRA-87-062113}.
Applications of the exotic quantum holonomy to quantum state
manipulation and adiabatic quantum
computation~\cite{Farhi-quant-ph-0001106} were also
proposed~\cite{%
Tanaka-PRL-98-160407,Miyamoto-PRA-76-042115,Tanaka-PRA-81-022320}.

Quantum maps describing 
periodically kicked 
systems~\cite{Berry-AP-122-26} 
are
useful to investigate the exotic quantum holonomy.  An adiabatic cycle
for a family of 
the
quantum maps induces a permutation among 
quasienergies,
which are determined by the eigenvalues of Floquet 
operators~\cite{Zeldovich-JETP-24-1006},
instead of 
their
eigenenergies~\cite{Tanaka-PRL-98-160407}.
We believe the experimental verification of the exotic quantum holonomy is now feasible considering the recent development of 
technology~\cite{Sadgrove-PRL-99-043002,Chabe-PRL-101-255702,Chaudhry-Nature-461-768,Morello-Nature-467-687}.
The exotic quantum holonomy in 
autonomous 
systems, 
in contrast
to periodically driven systems, requires either the 
divergence of the 
eigenenergies~\cite{Cheon-PLA-248-285} or
the exact 
crossing of eigenenergies~\cite{Cheon-PLA-374-144}.

It was recently pointed out that the exotic quantum holonomy is
closely associated with
Kato's exceptional points
(EPs)
located 
at the complexified parameter space
outside the adiabatic cycle%
~\cite{KatoExceptionalPoint}.
An EP is a degeneracy of non-Hermitian square matrix. Two eigenvalues
and eigenvectors of the $2\times2$ non-Hermitian matrix are
interchanged after parametric evolution along the cycle encircling the
EP, which resembles the holonomic behavior of the exotic quantum
holonomy.
However, the time evolution along the 
cycle
in the adiabatic limit generally do not induce the ``flip'' of
stationary states due to the presence of the the decay process, which
is inherent to non-Hermitian
systems~\cite{Uzdin-JPA-44-435302,Berry-JPA-44-435303,Uzdin-PRA-85-031804}.
On the other hand, if 
a
Hermitian cycle can be smoothly shrunk to the
non-Hermitian 
one so that the time spending in encircling the cycle is short enough, 
such an interchange of eigenspace 
can take place
within adiabatic time evolution.  
An example is a family of quantum kicked
spin-$\frac{1}{2}$~\cite{Kim-PLA-374-1958}, where the
interchange
between two eigenspaces in two level system due to the exotic quantum
holonomy correspond to the EP that resides in the complexified
parameter space of the quantum kicked spin.

As far as many interacting levels are concerned, situation becomes
much complicated since it is possible to find multiple
degeneracies. However, 
it is rather easier to deal with the case
that the adiabatic cycle
encircles several EPs associated with only two levels. It means that
the cycle contains many EPs, but each EP is doubly degenerate. This
has been recently studied in
Ref.~\cite{Ryu-PRA-85-042101,Lee-PRA-85-064103} by using $3\times3$
non-Hermitian matrix. 
The 
exotic quantum holonomy 
associated with multiple 
EPs
is
also studied in two-body Lieb-Liniger
model~\cite{Tanaka-JPA-46-315302}. However, the multiply degenerated
EP has been rarely investigated, see 
Ref.~\cite{Heiss-JPA-41-244010}
for triple 
EPs, and Ref.~\cite{Graefe-JPA-41-255206} for higher-order EPs.

In this paper we show highly degenerate 
EPs
can be systematically
constructed by using the 
quantum kicked top with appropriate
parameters chosen. 
The 
degeneracy of the EPs is given as
$2J+1$, where $J$ is the magnitude 
of angular momentum $J$ of the top.
We 
show
that the exotic quantum holonomy of the kicked quantum top
is intimately related to
the highly degenerate EP.

The plan of this manuscript is the following. In Sec.~\ref{sec:Ndim},
we introduce a quantum top under a rank-$1$ kick. We briefly explain
the consequence of the existence theorem of the exotic quantum
holonomy~\cite{Miyamoto-PRA-76-042115}. In Sec.~\ref{sec:Explicit}, we
show the presence of the exotic quantum holonomy using an
explicit expression of the solution of the eigenvalue problem, instead
of the existence theorem. This is possible only when the parameter of
the system satisfies a solvable condition. This condition
implies the presence of EPs with higher order, as shown in
Sec.~\ref{sec:dEP}. We 
show that the higher order EP is fragile
against perturbations
in
Sec.~\ref{sec:2EP}.
We also
explain the
correspondence between the exotic quantum holonomy and the 
remnants of
broken
higher order EPs. A summary 
is found in Sec.~\ref{sec:summary}.

\section{Quantum top under a rank-$1$ kick}
\label{sec:Ndim}

We introduce a quantum top (or spin) under a rank-$1$
kick~\cite{Combesqure-JSP-59-679} in this section.  We 
show the quasienergy and eigenspace anholonomies of this model with
the help of a theorem that ensures the existence of the exotic
quantum holonomy for quantum map under a rank-$1$
perturbation~\cite{Miyamoto-PRA-76-042115,Tanaka-PRL-98-160407}.

Let $\bvec{\hat{J}}$ denote the angular momentum of the top.
In the absence of kick, we suppose that the top rotates 
around
$z$-axis
with an angular frequency $\omega$. 
A rank-$1$ kick $\lambda \ketbra{v}{v}$, is applied periodically in time,
where $\lambda$ is the strength of the kick,
and $\ket{v}$ is a normalized vector. 
The time is normalized by the period of the 
kick.
The system is described by
the Hamiltonian
\begin{equation}
  \label{eq:Hamiltonian}
  \hat{H}(t) 
  = \omega \hat{J}_z 
  +\lambda\ketbra{v}{v} \sum_{n=-\infty}^{\infty} \delta (t -n)
  .
\end{equation}
We set $\hbar=1$ throughout this manuscript.
We assume that $\ket{v}$ belongs to 
$d$-dimensional
eigenspace of 
$\bvec{\hat{J}}^2=J(J+1)$, where $J$ is either an integer or 
half-integer, and $d=2J+1$.

The Hamiltonian (\ref{eq:Hamiltonian}) can be experimentally implemented 
by using
nuclear magnetic moment $\bvec{\hat{J}}$ under the influence of static magnetic field $\omega \hat{J}_z$ and the periodic kick $\ketbra{v}{v}$ 
composed by a polynomial of $\hat{J}_y$~\cite{Chose-PRA-78-042318,Chaudhry-Nature-461-768}.
The
polynomial depends on 
$J$; For example,
the rank-$1$ term for $J = \frac{1}{2}, 1$ and $\frac{3}{2}$ 
are
\begin{equation}
\begin{gathered}
  \hat{J}_y+\frac{1}{2}
  ,
  \qquad
  \frac{1}{2}(\hat{J}_y+1)\hat{J}_y 
  \\
  \text{and}\quad
  \frac{1}{6}\left(\hat{J}_y+\frac{3}{2}\right)
  \left(\hat{J}_y+\frac{1}{2}\right)
  \left(\hat{J}_y-\frac{1}{2}\right)
  ,
\end{gathered}
\end{equation}
respectively. 
These represent quadrupolar or higher order multipole
interactions of nuclear momentum.
We may replace $\hat{J}_y$ with $\bvec{\hat{J}}\cdot\bvec{n}$
for the
above
examples 
as long as 
a normalized vector
$\bvec{n}$ is not
parallel to $z$-axis.

We 
examine 
how the stationary states of the kicked top evolves when the kick strength 
$\lambda$ adiabatically varies.
Since the kicked top is 
a
periodically driven system, the stationary
states are the eigenvectors 
of the
Floquet 
operator describing
a unitary time evolution
during a unit time interval:
\begin{equation}
  \label{eq:floquetDef}
  \hat{U}(\lambda)
  \equiv e^{-i\omega \hat{J}_z}e^{-i\lambda \ketbra{v}{v}}
  .
\end{equation}
The 
real
parameter 
$\lambda$ is 
geometrically equivalent to a circle
because of the $2\pi$-periodicity of $\hat{U}(\lambda)$~\cite{Tanaka-PRL-98-160407},
which can be easily seen with the expansion 
\begin{equation}
  \hat{U}(\lambda)
  = e^{-i\omega \hat{J}_z}
  \left[(1-\ketbra{v}{v}) + \Lambda^{-1}\ketbra{v}{v}\right]
\end{equation}
with
\begin{equation}
  \label{eq:defalpha}
  \Lambda = e^{i\lambda}
  .
\end{equation}
Here
$\lambda$ 
runs from $0$ to 
$2\pi$ along a unit circle denoted by $C$.
Let $\ket{\varphi_n(\lambda)}$ ($n=0, \dots, d-1$) be an eigenvector of 
$\hat{U}(\lambda)$, i.e.,
\begin{equation}
  \hat{U}(\lambda)\ket{\varphi_n(\lambda)} 
  = z_n(\lambda)\ket{\varphi_n(\lambda)}
  ,
\end{equation}
where $z_n(\lambda)$ is the corresponding eigenvalue. Since $\hat{U}(\lambda)$
is unitary, $z_n(\lambda)$ 
lies
in the unit circle of the complex plane.
We introduce quasienergy of $E_n(\lambda)$ so as to satisfy
\begin{equation}
  \label{eq:defEn}
  z_n(\lambda) = e^{-iE_n(\lambda)}
  .
\end{equation}

According to 
the
theorem shown in 
Refs~\cite{Miyamoto-PRA-76-042115,Tanaka-PRL-98-160407}, 
the adiabatic cycle $C$ induces quasienergy and eigenspace
anholonomies, 
when two conditions are satisfied:
(1) $\exp(-i\omega \hat{J}_z)$ has no spectral degeneracy;
(2) all eigenvectors of $\exp(-i\omega \hat{J}_z)$ are neither
parallel nor orthogonal to $\ket{v}$.
The first 
is equivalent 
to a non-resonant condition
\begin{equation}
  \label{eq:nonresonnant}
  \omega \notin 
  \left\{\frac{2\pi{}q}{p}\;\Big|\;
    \text{$q$ and $p$ are integer and $0< |p| < d$}
  \right\}
  .
\end{equation}
The first condition 
together with
the assumption $\bracket{J,M}{v}\ne 0$
for all $M=-J,\dots,J$ implies the second condition, where
$\ket{J,M}$ is the standard basis of the angular momentum.
A typical example of $\ket{v}$ is 
\begin{equation}
  \label{eq:vdef}
  \ket{v}
  \equiv \frac{1}{\sqrt{d}}\sum_{M=-J}^{J}\ket{J,M}
  ,
\end{equation}
which will be employed 
below.

\begin{figure}
  \centering
  \includegraphics[width=0.35\textwidth]{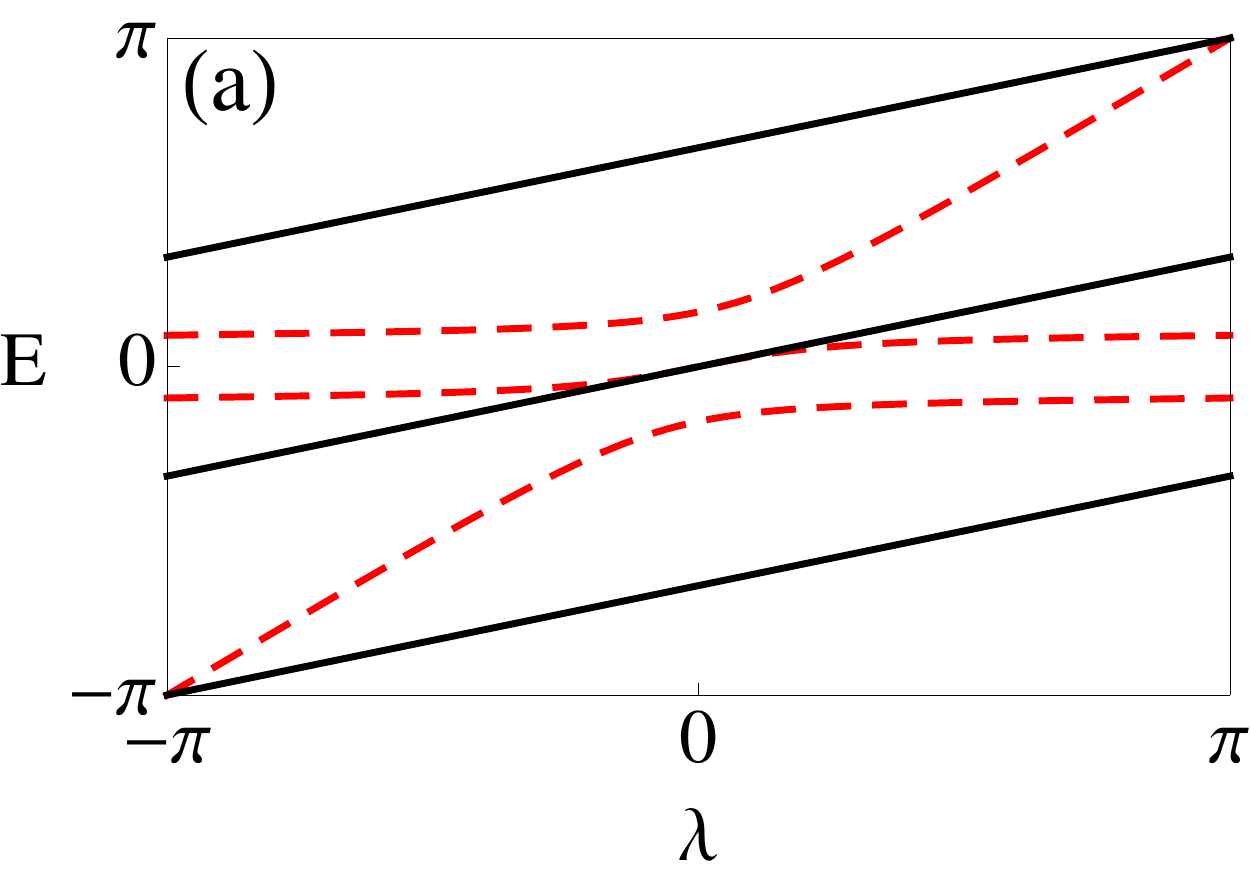}
  \\
  \includegraphics[width=0.35\textwidth]{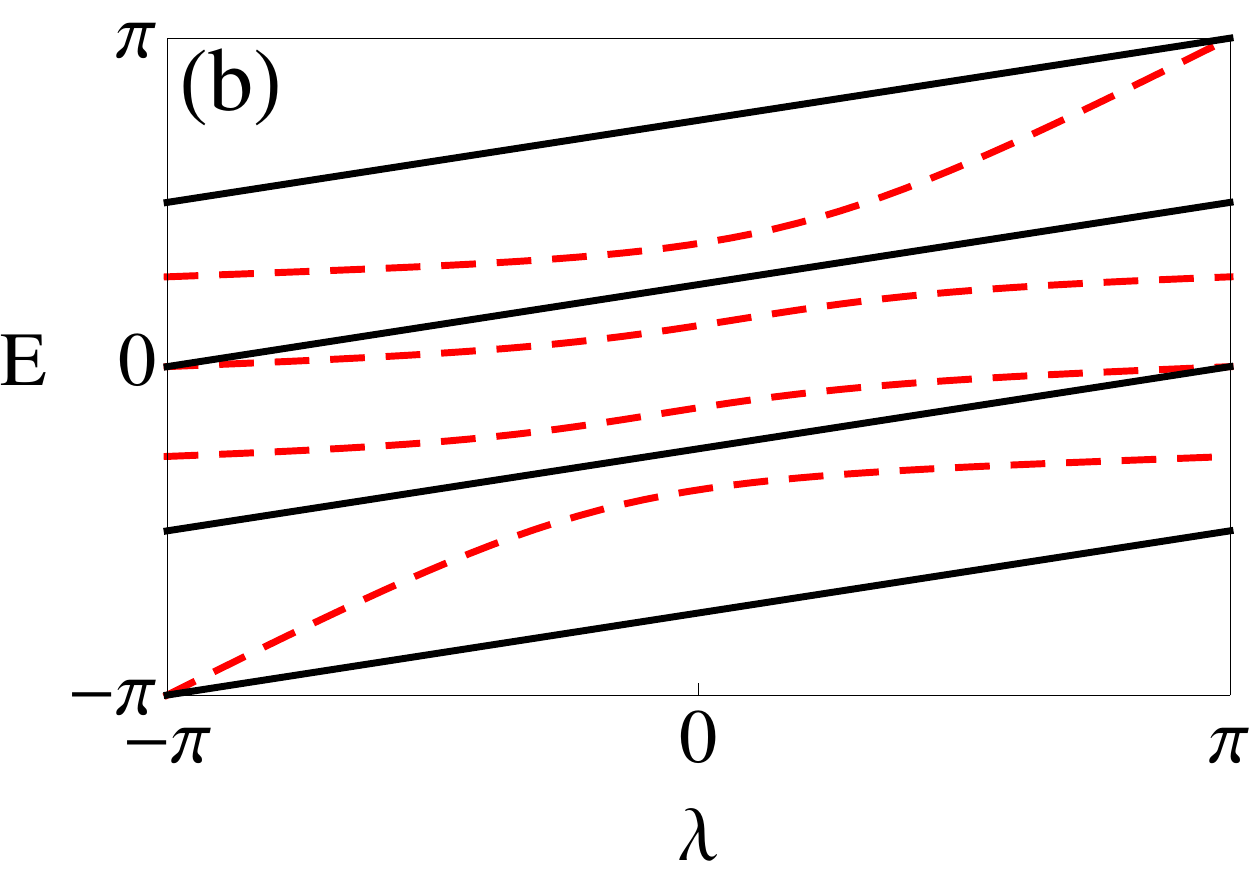}
  \caption{%
    Parametric evolution of quasienergies of the quantum top
    under a rank-$1$ kick~\eqref{eq:floquetDef} along the unitary cycle $C$.
    (a) $J=1$ case. Thick and dashed curves correspond to
    $\omega=2\pi/3$ and $\pi/6$, respectively. 
    (b) $J=3/2$ case. 
    Thick and dashed curves correspond to
    $\omega=\pi/2$ and $\pi/4$, respectively. 
    As shown later in the main text, 
    thick curves and dashed curves correspond to $(2J+1)$-EP cases
    (Sec.~\ref{sec:dEP}) and
    $2$EP cases (Sec.~\ref{sec:2EP}), respectively.
    }
  \label{fig:quasienergy}
\end{figure}

We now explain the permutation of quasienergies induced by the adiabatic 
cycle $C$.
We 
arrange
the quasienergies $\set{E_n(\lambda)}_{n=0}^{d-1}$ 
in the 
following order
\begin{align}
  \label{eq:orderedEn}
  0 \le E_0(\lambda) < E_1(\lambda) < \dots <E_{d-1}(\lambda) < 2\pi
\end{align}
at $\lambda=0$. 
The increment of $\lambda$ by $2\pi$ results in
\begin{align}
  \label{eq:Enanholonomy}
  E_n(\lambda + 2\pi) 
  = 
  \begin{cases}
      E_{n + 1}(\lambda)& \text{for $n = 0,\dots,d-2$}\\
      E_{0}(\lambda) + 2\pi & \text{for $n = d-1$}
  \end{cases}
  .
\end{align}
Thus
an adiabatic cycle $C$ 
increases
the quantum number 
by unity (with modulo $d$).
We emphasize that such an 
rearrangement
of
quasienergies occurs irrespective 
of $\omega$ and
$\ket{v}$, as long as both the non-resonant
condition~\eqref{eq:nonresonnant} and the cyclic condition are
satisfied.  
Several examples of the quasienergy anholonomy are
shown in Fig.~\ref{fig:quasienergy}.

The quasienergy anholonomy 
directly
implies 
the
anholonomy in eigenspaces.
Let us suppose that the state of the system is prepared 
in $\ket{J,M}$, which is a stationary state of the 
unperturbed Floquet operator 
$\hat{U}(\lambda)$
at $\lambda=0$.
Equation~\eqref{eq:Enanholonomy} then
implies that,
after the completion of the adiabatic cycle along $C$,
the state vector arrives 
at $\ket{J,s(M; C)}$ up to a phase factor,
where $s(M; C)$ is the eigenvalue of $\hat{J}_z$ of the final state.

In order to determine $s(M; C)$, we 
need
explicitly solve 
the eigenvalue problem of $\hat{U}(\lambda)$
at $\lambda=0$. It is straightforward to see that 
$\ket{J,M}$ ($M =-J,\dots,J$) is an eigenvector corresponding to 
a quasienergy $\omega M$.
Note that the mapping between the set of quasienergies
$\set{\omega M}_{M=-J}^J$ and $\set{E_n(\lambda)}_{n=0}^{d-1}$, 
which satisfies Eq.~\eqref{eq:orderedEn}
depends on $\omega$.
We 
explore it with some specific examples of
$J=\frac{1}{2}, 1$ and $\frac{3}{2}$
below.

First, we examine the case
of
$J=\frac{1}{2}$.
For example, in the interval $0 < \omega <2\pi$ where the non-resonant
condition~\eqref{eq:nonresonnant} is always satisfied, we have
\begin{align}
  E_0(0) = \frac{1}{2}\omega
  ,\quad
  E_1(0) = 2\pi -\frac{1}{2}\omega
  ,
\end{align}
which correspond to $M=\frac{1}{2}$ and $-\frac{1}{2}$, respectively.
According to Eq.~\eqref{eq:Enanholonomy}, $E_0(\lambda)$ arrives 
at $E_1(0)$ at $\lambda=2\pi$. Hence, the quantum state
initially prepared to be in 
$\ket{1/2, 1/2}$ 
at $\lambda=0$ is delivered
to 
$\ket{1/2, -1/2}$ 
after the completion of the adiabatic cycle $C$.
Similarly, 
the adiabatic cycle $C$ delivers 
$\ket{1/2, -1/2}$ 
to 
$\ket{1/2, 1/2}$. 
This implies 
$s(\pm\frac{1}{2};C)=\mp\frac{1}{2}$.

Second, we examine the case {$J=1$}.
We show that two intervals $0 < \omega < \pi$ 
and $\pi < \omega < 2\pi$ 
provide
different types of $s(M,C)$.
Note that the non-resonant condition~\eqref{eq:nonresonnant}
is always satisfied in both intervals.
In the former case, 
\begin{align}
  E_0(0) = 0
  ,\quad
  E_1(0) = \omega
  \quad
  \text{and}\quad
  E_2(0) = 2\pi - \omega
  ,
\end{align}
which correspond to $M=0$, $1$ and $-1$, respectively.
Using a similar argument applied to the $J=\frac{1}{2}$ case 
mentioned
above,
we find 
$s(0;C)=1$, $s(1;C)=-1$ and $s(-1;C)=0$, 
which comprise
a cyclic permutation.
Namely, the itinerary of $\ket{J,M}$ induced by 
the adiabatic cycle $C$ is
\begin{align}
  \ket{J,0}\mapsto\ket{J,1}\mapsto\ket{J,-1}
  .
\end{align}
For another interval $\pi < \omega < 2\pi$, we have
\begin{align}
  E_0(0) = 0
  ,\quad
  E_1(0) = 2\pi - \omega
  ,\quad
  E_2(0) = \omega
\end{align}
to 
satisfy
Eq.~\eqref{eq:orderedEn}. Hence the corresponding
adiabatic itinerary of $\ket{J,M}$ is
\begin{align}
  \ket{J,0}\mapsto\ket{J,-1}\mapsto\ket{J,1}
  .
\end{align}
Hence the itinerary is suddenly 
changed
around a critical point
$\omega=\pi$.
In this sense, we may choose the itinerary by varying
$\omega$.
Now it is straightforward to extend the present analysis 
to
arbitrary $J$ and $\omega$.

\section{Explicit expressions of eigenvalues and 
  eigenvectors}
\label{sec:Explicit}

So far we have explained the exotic quantum holonomy of
the kicked tops \eqref{eq:Hamiltonian} using the general theorem 
shown in Refs.~\cite{Tanaka-PRL-98-160407,Miyamoto-PRA-76-042115}.
In this section, we explain the details of the anholonomies 
with the help of the explicit expressions of quasienergy and 
eigenvectors for an arbitrary $J$. This is possible when we choose
a ``solvable'' value of $\omega$.
This also helps us to examine the complexification 
of $\lambda$, as shown in the following sections.

We here examine the case 
$\omega=2\pi/d$. Note that
the following argument is
also applicable to the case $\omega=2\pi r/d$ with an integer $0\le r < d$.
We 
assume that $\ket{v}$ 
satisfies
Eq.~\eqref{eq:vdef}.
This allows us to introduce a mapping of
the kicked top into a kicked particle in a periodic lattice
(or, a kicked tight-binding model).
We introduce basis vectors
\begin{equation}
  \label{eq:phiDef}
  \ket{m}
  \equiv \frac{1}{\sqrt{d}}\sum_{M=-J}^{J}e^{i 2\pi M m/d}\ket{J,M}
  ,
\end{equation}
which describes a localized state at ``$m$-th site''
for $0 \le m < d$. 
From the assumption $\omega=2\pi/d$, the unperturbed Floquet
operator $\hat{U}(0)$ is 
\begin{equation}
  \hat{U}(0)
  = \sum_{m=0}^{d-1}\ketbra{{m-1}}{{m}}
  ,
\end{equation}
which describes a non-dispersive 
motion of
a
particle in a 
one-dimensional
periodic lattice.
Since we have chosen $\ket{v} = \ket{m=0}$
(see Eq.~\eqref{eq:vdef}),
we obtain the Floquet operator in the $\ket{m}$-representation:
\begin{equation}
  \label{eq:UTB}
  \hat{U}(\lambda)
  = \sum_{m=1}^{d-1}\ketbra{{m-1}}{m}
  +\Lambda^{-1}\ketbra{{d-1}}{0}
  .
\end{equation}
This implies that an extra phase factor 
$\Lambda^{-1}$ is added along a ``hopping'' from $0$-th to $(d-1)$-th site.
Equation~\eqref{eq:UTB} is represented in the matrix representation 
with basis vectors 
$\set{\ket{m}}_{m=0}^{d-1}$ as follows:
\begin{equation}
  \label{eq:Umatrix}
  {U}(\lambda)
  =
  \begin{bmatrix}
    0&1\\
     &0&1\\
     & &\ddots&\ddots&\\
    0& &      &\ddots&1\\
    \Lambda^{-1}&0 &      &      &0
  \end{bmatrix}
  .
\end{equation}
It is straightforward to obtain the characteristic 
equation
of 
$\hat{U}(\lambda)$:
\begin{align}
  \label{eq:charU}
  \det\{z-{U}(\lambda)\}
  &
  = z^d -  \Lambda^{-1}
  = 0,
\end{align}
whose solution is
\begin{equation}
  \label{eq:zMdef}
  z_M(\lambda) = e^{-i (2\pi M + \lambda)/d}
  ,
\end{equation}
and the corresponding quasienergy is 
\begin{equation}
  \label{eq:EMdef}
  E_M(\lambda) 
  = \frac{\lambda + 2\pi M}{d}
  ,
\end{equation}
with
$M=-J,\dots, J$.
We also find the corresponding normalized eigenvector
\begin{equation}
  \ket{\xi_M(\lambda)}
  = \frac{1}{\sqrt{d}}\sum_{n=0}^{d-1}e^{-i n E_M(\lambda)} \ket{n}
  .
\end{equation}

In terms of basis $\ket{\xi^{}_M(\lambda)}$, the eigenspace 
holonomy is
regarded as
an increment of quantum number $M$, i.e.,
\begin{align}
  \left|\bracket{\xi^{}_{M'}(\lambda+2\pi)}{\xi^{}_M(\lambda)}\right|^2
  = \delta_{M', M+1 \mod d}
  .
\end{align}
On the other hand, we need to identify which $\ket{J,M}$ is
parallel to $\ket{\xi^{}_{M'}(0)}$ for a given $M'$ to 
completely understand
the eigenspace anholonomy. 
We show that $\ket{\xi^{}_M(0)}$ is parallel
to $\ket{J,M'}$ only 
if
$M - M' = 0 \mod d$, i.e.,
\begin{align}
  \bracket{J,M'}{\xi^{}_M(0)}
  &
  = \delta_{M, M' \mod d}
  .
\end{align}
We emphasize that this is applicable 
to any
arbitrary $J$.

\section{Higher order exceptional points 
  behind exotic quantum holonomy}
\label{sec:dEP}

In this section, we examine the EPs 
in the kicked quantum tops~\eqref{eq:Hamiltonian} 
under the specific choice of parameter $\omega=2\pi/d$, which is
examined in the previous section.
We will show that the degree 
or the multiplicity
of the EP is 
$d$, the highest possible value of the degree.
We denote such
an EP as $d$EP to distinguish it
from
conventional EPs whose order is $2$.

So far, we have assumed that $\lambda$ is real, or equivalently,
$|\Lambda|$ is unity (see, Eq.~\eqref{eq:defalpha}). 
From now
we complexify $\lambda$ to investigate EPs.
This makes $\hat{U}(\lambda)$ a non-unitary operator, which 
may be regarded as an effective time evolution 
operator that describes conditional measurement 
processes~\cite{Srinivas-OptAc-28-981,Mohseni-JCP-129-174106}.
The expression of the eigenvalues (Eq.~\eqref{eq:zMdef}) remains 
intact regardless of
the complexification of $\lambda$.
On the other hand, because $\hat{U}(\lambda)$ is 
no longer
unitary
when $\lambda$ is not real, 
the left and the right eigenvectors of $\hat{U}(\lambda)$
become 
different so that the left eigenvector is separately introduced as 
$\bra{\xi_M^{\mathrm{L}}(\lambda)}$~\cite{biorthogonal}.
Both eigenvectors read
\begin{equation}
  \begin{split}
  \ket{\xi^{}_M(\lambda)}
  &
  = \frac{1}{\sqrt{d}}\sum_{m=0}^{d-1}\{z_M(\lambda)\}^m \ket{m}
  ,
  \\
  \bra{\xi_M^{\mathrm{L}}(\lambda)}
  &
  = \frac{1}{\sqrt{d}}\sum_{m=0}^{d-1}\{z_M(\lambda)\}^{-m} \bra{m}
  ,
  \end{split}
\end{equation}
which satisfy
$\bracket{\xi_{M'}^{\mathrm{L}}(\lambda)}{\xi^{}_M(\lambda)}
=\delta_{M'M}$.

The spectral degeneracy of $\hat{U}(\lambda)$ occurs 
when
$\rmIm\lambda = -\infty$, where all eigenvalues 
accumulate
at $z=0$.
This can be 
easily
understood from the matrix representation
${U}(\lambda)$ (Eq.~\eqref{eq:Umatrix}),
which
converges to the $d\times d$ nilpotent Jordan 
block in the limit $\rmIm\lambda\to -\infty$:
\begin{align}
  {U}(\lambda)
  \to
  \begin{bmatrix}
    0&1\\
     &0&1\\
     & &\ddots&\ddots&\\
    & &      &\ddots&1\\
    0& &      &      &0
  \end{bmatrix}
  .
\end{align}
Hence we conclude that 
$\Lambda=\infty$
is an EP
of the order $d$.
Also, we find that, from the characteristic equation (Eq.~\eqref{eq:charU}),
there is 
another
$d$EP at $\Lambda=0$, where the eigenvalues of
$\hat{U}(\lambda)$ accumulate at $z=\infty$.

\begin{figure}
  \centering
  \includegraphics[width=0.35\textwidth]{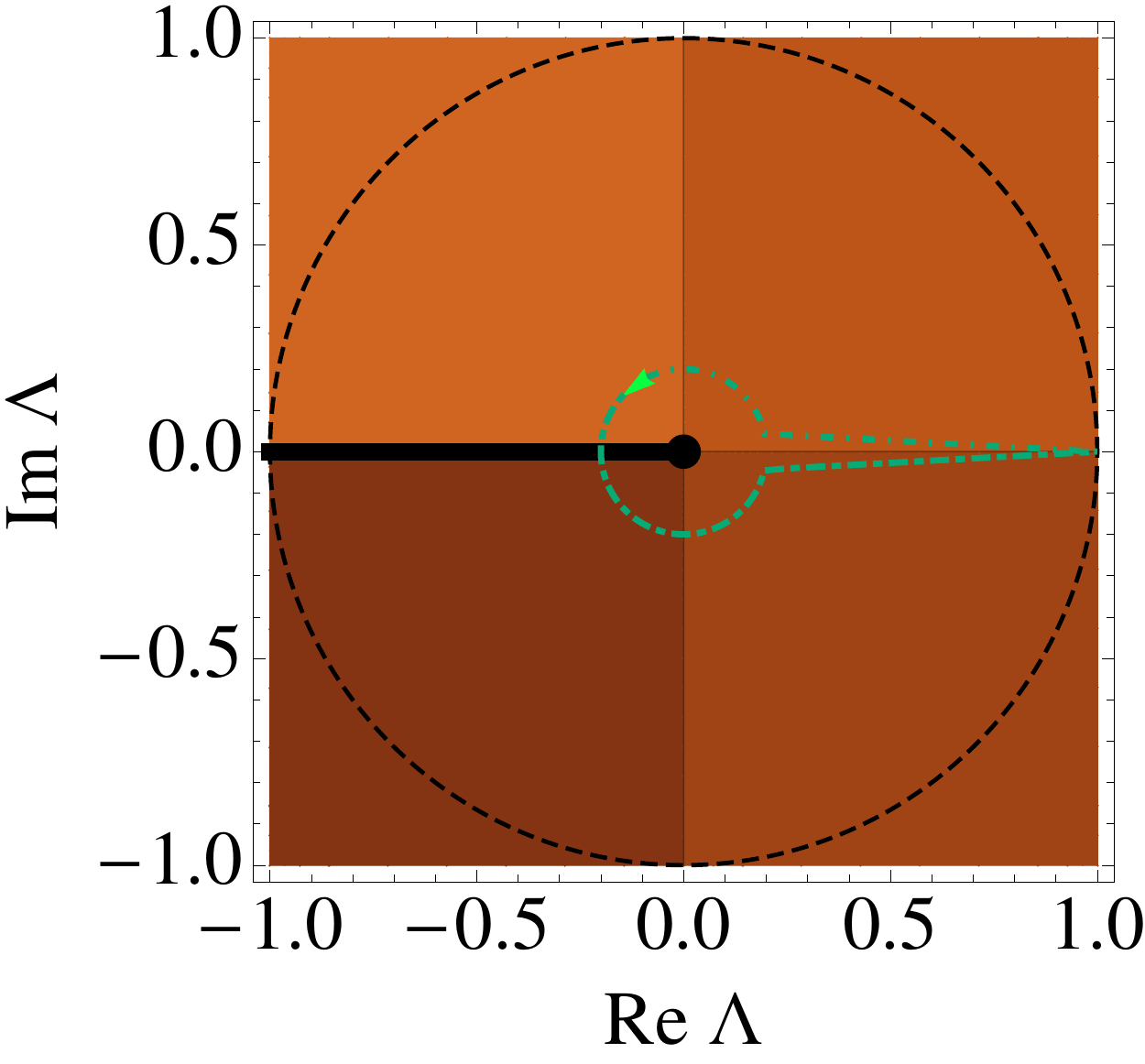}
  \caption{%
    (Color online)
    A contour plot of $\rmRe E_{-1}$ in the $\Lambda$-plane ($\Lambda$ is
    defined in Eq.~\eqref{eq:defalpha}).
    Lighter (darker) color indicates large (smaller) value of $\rmRe E_{-1}$.
    Parameters are $J=1$ and $\omega=2\pi/(2J+1)$. There is a $3$EP at 
    $\Lambda=0$, which is indicated by filled circle. A branch 
    cut, which
    emanates 
    from the 
    $3$EP,
    is drawn by a bold line. 
    A dash-dotted curve $C'$ is a smooth deformation of $C$.
    The exotic quantum holonomy induced by $C$ can be emulated by
    the non-Hermitian cycle $C'$.
    }
  \label{fig:EP3}
\end{figure}

An EP is the branch point of eigenvalues. We choose 
the line from the $d$EP at the origin to 
$-\infty$ 
in $\Lambda$
plane as the branch cut represented by the thick horizontal line in
Fig.~\ref{fig:EP3}.
This choice is consistent 
with the 
analytic continuation of 
$z_M(\lambda)$ considered below.
We start from the unit circle in the $\Lambda$-plane, where
$\lambda$ is real-valued. Because of the presence of the eigenvalue
anholonomy, we need to introduce a discontinuous point of $z_M(\lambda)$
in the unit circle of $\Lambda$. 
Here we suppose 
$z_M(\pi+0) = z_{M+1 {\rm \ mod\ }1}(\pi-0)$. 
Hence $z_M(\lambda)$ is discontinuous at $\Lambda=-1$ in $C$.
For each point in the unit circle of $\Lambda$, 
we extend $z_M(\lambda)$ along 
the radial direction in the $\Lambda$-plane.
This uniquely specifies $z_M(\lambda)$ 
in the whole $\Lambda$-plane. 

The 
variation of $\Lambda$ along $C$ (Fig~\ref{fig:EP3})
induces permutation of the quasienergies
\begin{equation}
\label{eq:J1permutation}
E_{M=-1}\mapsto E_{M=0}\mapsto E_{M=1}  
. 
\end{equation}
This is 
an extension of 
the EP-interpretation of the quasienergy
anholonomy 
originally introduced in Ref.~\cite{Kim-PLA-374-1958} to a family of 
multiple-level systems.

We explain an emulation of the exotic quantum holonomy with EPs by
deforming the unitary cycle $C$ into non-Hermitian cycles, say, $C'$.
Suppose $C'$ enclose the $d$EP and connect between the $d$EP
and the initial point of the cycle $C$. This is depicted in 
Fig.~\ref{fig:EP3}, as for $d=3$ case.
Since the change of eigenvectors essentially occurs along the small cycle
around the $d$EP, 
we may say that only the contribution from the $d$EP is relevant.

\section{Correspondence between 
  the exotic quantum holonomy and 
  the fragments of $d$EP}
\label{sec:2EP}

Even a 
small perturbation can destroy
a
$d$EP ($d>2$).
In this section, we examine the stability of the $d$EPs 
of the quantum kicked top~\eqref{eq:floquetDef} at $\omega=2\pi/d$,
for $d=3$ and $4$.
We 
show that the $d$EP is broken into $2$EPs,
the number of which is $d-1$,
when we slightly vary $\omega$ from $2\pi/d$.
In contrast to such a catastrophe of the $d$EPs, the exotic quantum holonomy
is stable against such small perturbations.
This raises another question how the correspondence between the exotic 
quantum holonomy and the fragments of the $d$EP is established, which 
is also examined in this section.

\begin{figure}
  \centering
  \includegraphics[width=0.4\textwidth]{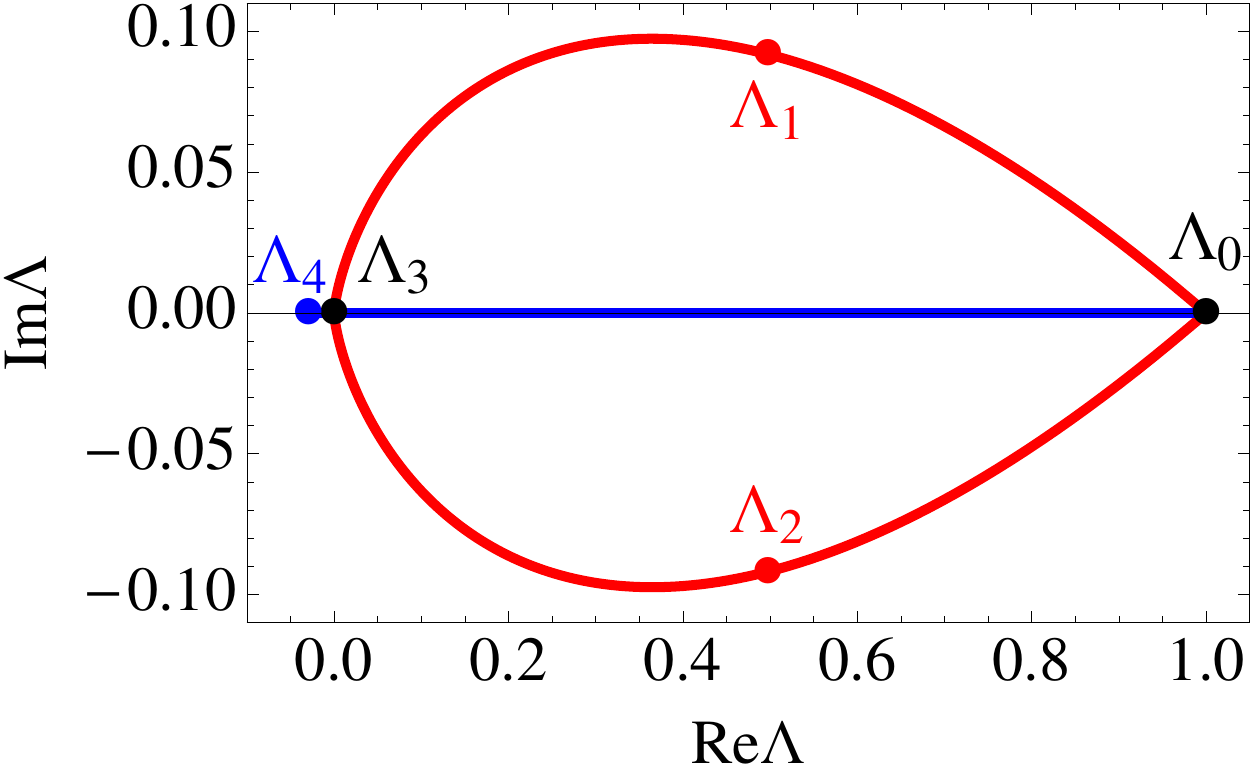}    
  \caption{(Color online) 
    Parametric evolution of EPs of the quantum kicked top $J=1$
    within the unitary cycle $C$ in the $\Lambda$-plane. 
    $\omega$ is varied within the interval $[0,\pi]$.
    There is a triple degeneracy point $\Lambda_0$($=1$) at $\omega=0$.
    Note that the system is unitary (diagonalizable) at $\Lambda=\Lambda_0$.
    As $\omega$ is increased in the interval $(0,2\pi/3)$, two $2$EPs
    evolve along the curves that pass $\Lambda_1$ and $\Lambda_2$
    at 
    $\omega=\pi/6$. 
    These $2$EPs 
    merge 
    at $\omega=2\pi/3$ to form 
    a $3$EP, which locates at $\Lambda_3$($=0$). If 
    $\omega$ is increased further ($2\pi/3 < \omega < \pi$), 
    there are two $2$EPs, which are fragments of the 
    $3$EP $\Lambda_3$,
    in the real axis.
    Finally, at $\omega=\pi$, a $2$EP arrives at $\Lambda_0$
    to become a diabolic point,
    and the other arrives at $\Lambda_4$($<0$).
  }
  \label{fig:lin00_3_EP}
\end{figure}

We show a detailed analysis of $J=1$ 
(i.e., $d=3$)
case.
In Fig.~\ref{fig:lin00_3_EP}, we depict 
numerically obtained
$\omega$-dependence 
of 
EPs.
At $\omega=0$, we have a triple degeneracy point at $\Lambda=1$,
marked as $\Lambda_0$,
where the Floquet operator is unitary.
In the interval $0<\omega < 2\pi/3$, there are two $2$EPs within
the unit circle $\left|\Lambda\right|<1$
(e.g., $\Lambda_1$ and $\Lambda_2$ in Fig.~\ref{fig:lin00_3_EP}).
These two $2$EPs merge to form 
a $3$EP 
($\Lambda_3$ in the figure)
at $\omega=2\pi/3$.
When $\omega$ is increased further,
the $3$EP again split into two $2$EPs.
One of them evolves along the positive real axis, and
finally arrives at $\Lambda=0$ at $\omega=\pi$.
The other 
$2$EP evolves along the negative real axis, and
arrives at 
$\Lambda_4$.
We note that it
suffices
to examine the interval 
$0\le \omega\le \pi$ due to
a reflection symmetry about $\omega=\pi$ in the 
$\omega$-dependence of the configuration of EPs.

We provide an analytic argument to the 
above numerical findings.
We examine the characteristic polynomial of 
the Floquet 
operator~\eqref{eq:floquetDef} given as
\begin{align}
  f(z) 
  &
  \equiv
  \det\left|z - \hat{U}(\lambda)\right|
  \nonumber \\ &
  =
  z^3 + f_2 z^2 + f_1 z + f_0,
\end{align}
where 
$f_2 \equiv \left(2+\Lambda^{-1}\right)\mu$,
$f_1 \equiv - \left(1+2\Lambda^{-1}\right)\mu$,
$f_0 \equiv-\Lambda^{-1}$
and
\begin{align}
  \mu 
  &
  \equiv - \frac{1}{3}(1+2\cos\omega)
  .
\end{align}
Note that $-1 \le \mu \le \frac{1}{3}$ holds for real $\omega$.
Following the standard prescription to solve cubic equations~\cite{Cardano}, 
we introduce $p$ and $q$:
\begin{gather}
  p 
  \equiv\frac{3 f_1 - f_2^2}{9}
  ,\quad
  q
  \equiv 
  \frac{2 f_2^2 - 9 f_2 f_1 + 27 f_0}{27}
  ,
\end{gather}
from which the discriminant is defined
as
\begin{align}
  \label{eq:D_def}
  D 
  \equiv -27(q^2+4p^3)
  .
\end{align}
The presence of 
spectral
degeneracy in $\hat{U}(\lambda)$ is equivalent
to
the condition $D=0$. Indeed, when $D=0$ holds, the 
solutions
of the characteristic equation $f(z)=0$ are,
according to the Cardano formula~\cite{Cardano}, 
\begin{align}
  z =
  z_{\rm c}+2\left(-\frac{q}{2}\right)^{1/3}
  \quad\text{and}\quad
  z_{\rm c}-\left(-\frac{q}{2}\right)^{1/3}
  ,
\end{align}
where 
$z_{\rm c} \equiv -f_2/{3}$. The latter solution is
doubly degenerated.
Hence, the spectrum of $\hat{U}(\lambda)$ is triply degenerate if and 
only if 
both
$D=0$ and $q=0$ 
hold.

We show that there are, at most, four spectral degeneracy points, which are
either diabolical or exceptional point, in the $\Lambda$-plane. 
This is because the discriminant $D$ (Eq.~\eqref{eq:D_def}) is
a fourth order polynomial in $\Lambda^{-1}$:
\begin{align}
  \label{eq:D_polynomial}
  D 
  =
  \sum_{n=0}^4 D_n \left(\Lambda^{-1}\right)^{4-n}
  ,
\end{align}
with
$D_0=D_4=4(\mu +1)\mu^3$,
$D_1=D_3=4(\mu +1)(9+5\mu)\mu^2$
and
$D_2=-3(\mu +1)(9-9\mu-21\mu^2-11\mu^3)$.

We also show that degeneracy points form a ``conjugate pair'':
If
$\Lambda$ is 
a degeneracy point, 
then so is $\Lambda^{-1}$.
The reason is that
$D=0$ is a reciprocal equation, whose 
coefficients satisfy symmetry relations $D_0=D_4$ and $D_1=D_3$.

We 
now
show that the triple 
degeneracy, obtained by solving $D=0$ and $q=0$,
occurs 
at
$(\omega,\Lambda) = (0, 1)$ 
and 
$(2\pi/3, 0)$, 
as 
shown
in 
Fig.~\ref{fig:lin00_3_EP}.
$q$ is expressed as a polynomial of
$\Lambda^{-1}$:
\begin{align}
  q
  = 
  \sum_{n=0}^3 q_n \left(\Lambda^{-1}\right)^{3-n}
  ,
\end{align}
with
$q_0=2\mu^3/27$, $q_1=2\mu^2(3+2\mu)/9$, 
$q_2=-(9-15 \mu^2-8 \mu^3)/9$
and
$q_3=2\mu^2(9+8\mu)/27$.
$D$ and $q$
vanish simultaneously if and only if the resultant
(Sylvester's determinant) $R(D,q)$ of these
polynomials vanish~\cite{Resultant}:
\begin{align}
  R(D,q)
  &
  =
  \left|
  \begin{matrix}
    D_0& D_1& D_2& D_3& D_4 &0&0&0\\
    0  & D_0& D_1& D_2& D_3& D_4 &0&0\\
    0  & 0  & D_0& D_1& D_2& D_3& D_4 &0\\
    0  & 0  & 0  & D_0& D_1& D_2& D_3& D_4\\
    q_0& q_1& q_2& q_3 &0&0&0&0\\
    0  & q_0& q_1& q_2& q_3 &0&0&0\\
    0  & 0  & q_0& q_1& q_2& q_3 &0&0\\
    0  & 0  & 0  & q_0& q_1& q_2&q_3&0\\
    0  & 0  & 0  & 0  & q_0& q_1& q_2&q_3\\
  \end{matrix}
  \right|
  \\ &
  = 
  \frac{16}{27} (3 - \mu)^3 \mu^9 (1 + \mu)^9
  .
\end{align}
The resultant $R(D,q)$ vanish only when 
$\mu$ equals to 
either $0$ or $-1$.
This is the 
condition of the triple degeneracy.
These two cases (T1) and (T2) are examined in the following.
\paragraph*{Case (T1) $\omega=\frac{2\pi}{3}$ $(\mu=0)$:} 
As shown in Sec.~\ref{sec:Explicit},
the characteristic polynomial is $f(z) = z^3 - \Lambda^{-1}$.
Hence 
a
$3$EP locates at 
$\Lambda=\infty$, whose conjugate pair $\Lambda=0$ is also a $3$EP.
\paragraph*{Case (T2) $\omega=0$ $(\mu=-1)$:} 
Because the characteristic polynomial is $f(z) = (z-1)^2(z-\Lambda^{-1})$,
the Floquet operator has eigenvalues $1$ and $\Lambda^{-1}$, 
former of which
is doubly degenerate, and corresponding quasienergy is real.
The degree of the ``Hermitian'' degeneracy at $\Lambda=1$,
which corresponds to $\Lambda_0$ in Fig.~\ref{fig:lin00_3_EP},
is $3$. Otherwise the degree is $2$.

In the following, we examine 
the cases (D1)--(D3) where there 
are,
at most, doubly spectral degeneracies.
\paragraph*{Case (D1) $\omega=\pi$ $(\mu = \frac{1}{3})$:} 
There are 
three
doubly
degeneracy points
$\Lambda=\Lambda_{\pm}$ and $1$, 
which are 
the
solutions of $D=0$.
We have a conjugate pair of 2EPs 
$\Lambda_{\pm}= -(17 + 12 \sqrt{2})^{\pm 1}$, which satisfy 
$\Lambda_{+}\Lambda_{-}=1$.
Note that $\Lambda_-$ is 
located inside
the unit circle, and corresponds 
to $\Lambda_4$ in Fig.~\ref{fig:lin00_3_EP}.
Also, there are a Hermitian degeneracy point at $\Lambda=1$.
\paragraph*{Case (D2) $\frac{2\pi}{3} < \omega < \pi$ 
  $(0 < \mu < \frac{1}{3})$:} 
We
explain that
a pair of EPs 
lies
in the positive real axis, 
and the other pair 
lies
in the negative real axis.
The solution of $D=0$ of
\begin{align}
  {D}
  = \frac{4\mu ^3(\mu +1)}{\Lambda^{2}}
  \left[
  \left(y+\frac{9+5 \mu}{2\mu}\right)^2
  -\frac{27(1+\mu)^2}{4\mu^3}
  \right]
\end{align}
is given as
\begin{align}
  y_{\pm}
  \equiv 
  \frac{1}{2\eta^3}\left[-(3+5\eta^2)\eta \pm (1+3\eta^2)\right]
  ,
\end{align}
with
\begin{align}
  \label{eq:def_y}
  y = \Lambda + \Lambda^{-1}
\end{align}
and
\begin{align}
  \label{eq:def_eta}
  \eta\equiv \sqrt{\frac{|\mu|}{3}}
  .
\end{align}
Note that $0 < \eta < 1/3$ holds for $0<\mu < 1/3$.
We solve Eq.~\eqref{eq:def_y} to obtain $\Lambda$. Let us examine the
case $y=y_+$ first. It is straightforward to see that the solutions are
\begin{align}
  \Lambda_{\pm}^{(0)}
  \equiv \frac{1}{2}\left(y_+ \pm \sqrt{(y_+)^2-4}\right)
  ,
\end{align}
which 
form
a conjugate pair. We show that $\Lambda_{\pm}^{(0)}$ are
positive and real.
First, we examine the discriminant of the quadratic equation
\begin{align}
  (y_+)^2-4
  = \frac{1}{4\eta^6}\left(1 -\eta\right)^3(1+3\eta^2)(1-3\eta)
  ,
\end{align}
which is positive for $0 < \eta < 1/3$.
Second, we examine the sign of $y_{+}$. 
We therefore have
$|y_{+}| > 2$
for $0 < \eta < 1/3$. Hence the sign of $y_{+}$ is independent
of $\eta\in (0,1/3)$. As 
$
\eta
\to 0+
$, 
it is easy to see $y_+>0$.
This implies that $y_+$ is positive for $\eta\in (0,1/3)$.
Hence $\Lambda_{\pm}^{(0)}$ are also real and positive.

In a similar way, we examine the case 
$y=y_-$, where we have
\begin{align}
  \label{eq:x_+}
  \Lambda_{\pm}^{(1)}
  \equiv \frac{1}{2}\left(y_- \pm \sqrt{(y_-)^2-4}\right)
  ,
\end{align}
which also 
form
a conjugate pair. 
It is straightforward to
see $\Lambda_{\pm}^{(1)}$ are real and negative because 
$y_- < 0$ and $(y_-)^2 - 4>0$ hold.

\paragraph*{Case (D3) $0 < \omega < \frac{2\pi}{3}$ 
  $(-1 < \mu < 0)$:} 
We explain that there are two conjugate pair of 2EPs 
in the $\Lambda$-plane, which can be proved in a similar way above.
The difference from the 
above case 
is that 
these EPs are not in the real axis.
As is seen in Fig.~3, the trajectories (depicted by red curve) of two 
$2$EPs
within the unit circle correspond to this case.

We solve $D=0$ to obtain $y$ (see Eq.~\eqref{eq:def_y}). The solutions
are
\begin{align}
  y_{c}
  \equiv 
  \frac{1}{2\eta^3}\left[(3-5\eta^2)\eta + i(1-3\eta^2)\right]
  .
\end{align}
and $y_{c}^*$,
where $\eta$ is defined in Eq.~\eqref{eq:def_eta}. 
Note that $0 < \eta < 1/\sqrt{3}$ holds for $-1<\mu < 0$.
For $y=y_{c}$, the solution of Eq.~\eqref{eq:def_y} are
\begin{align}
  \Lambda_{\pm}^{(c)}
  \equiv \frac{1}{2}\left(y_c \pm \sqrt{(y_c)^2-4}\right)
  ,
\end{align}
which form a conjugate pair of degeneracy point.
We note that $[\Lambda_{\pm}^{(c)}]^*$ are also EPs.
The discriminant of the quadratic equation is
\begin{align}
  &
  (y_{c})^2-4
  \nonumber\\ &
  = \frac{\left[3\eta(1-\eta^2/3)+ i(1-3\eta^2)\right]
  (1-3\eta^2)(3\eta+i)}{4\eta^6}
  ,
\end{align}
which is nonzero and complex-valued as long as $-1 < \mu < 0$.
Hence $\Lambda_{\pm}^{(c)}$ and $[\Lambda_{\pm}^{(c)}]^*$ 
are not in the real axis.

We note that 
the
3EP is fragile against perturbations 
of
$\mu$. 
Indeed, once $\mu$ is varied from $0$, the resultant $R(D,q)$ becomes nonzero, 
which implies the absence of triple degeneracy.

We proceed to examine the correspondence between the exotic quantum holonomy
and the fragments of the $3$EPs.
To establish this, we
explain the emulation of the exotic quantum holonomy with EPs by
deforming the unitary cycle $C$ into non-Hermitian cycles, say, $C'$.
As for the $3$EP case ($\omega=2\pi/3$), we refer the previous section
and Fig~\ref{fig:EP3}. Because there is only a single EP, the resultant
Riemann surface structure is rather simple.

Due to the rupture of $3$EP into several EPs, the Riemann surface
structure becomes complicated. 
We here examine the case $\omega=\pi/6$, where the eigenspace anholonomy 
is 
equivalent to
the case $\omega=2\pi/3$ 
(see Fig.~\ref{fig:quasienergy}). There are two $2$EPs
$\Lambda_1$ and $\Lambda_2$
(Fig.~\ref{fig:lin00_3_EP}). 

\begin{figure}
  \centering

  \includegraphics[width=0.23\textwidth]{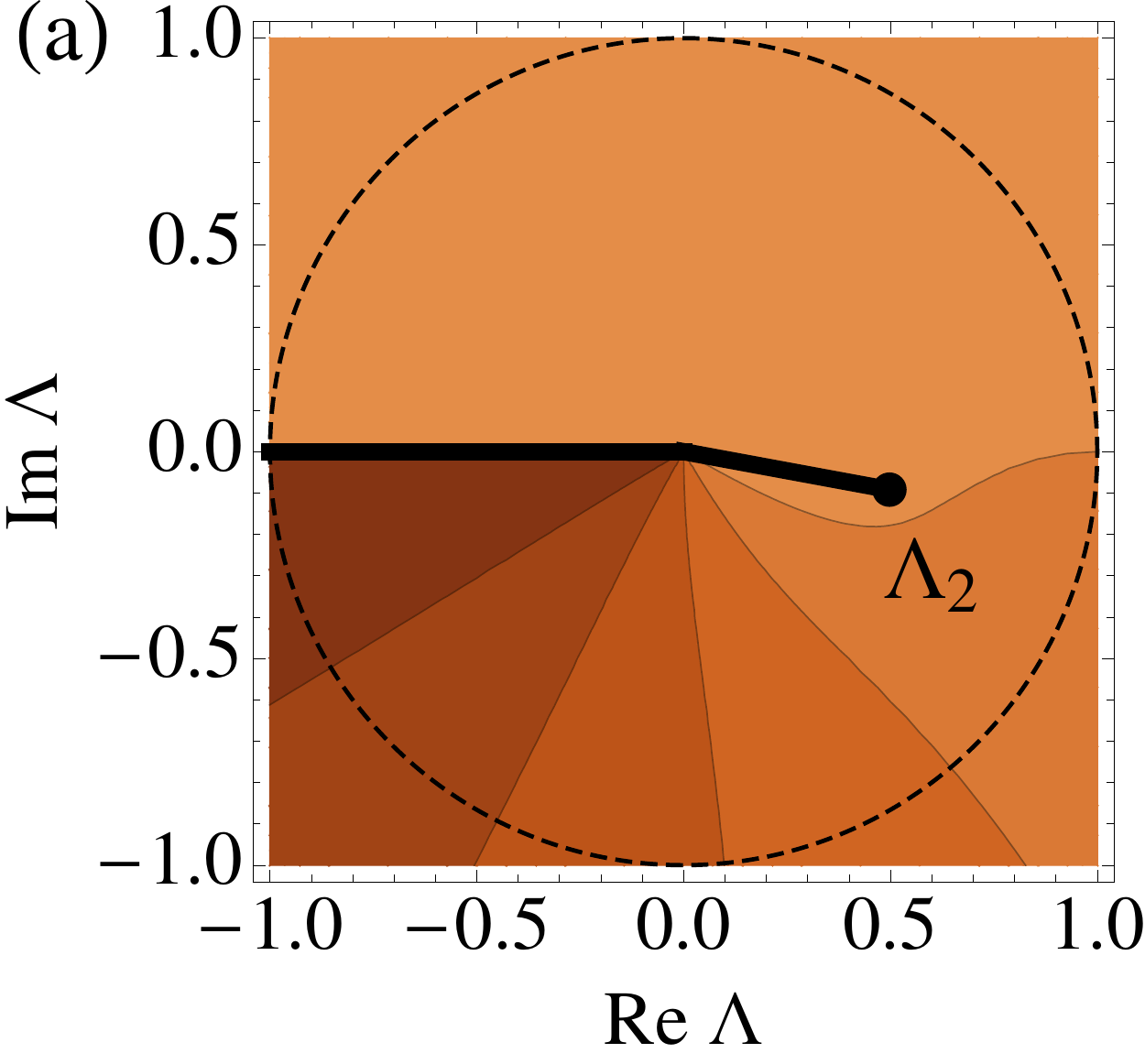}
  \hspace{0.5em}
  \includegraphics[width=0.23\textwidth]{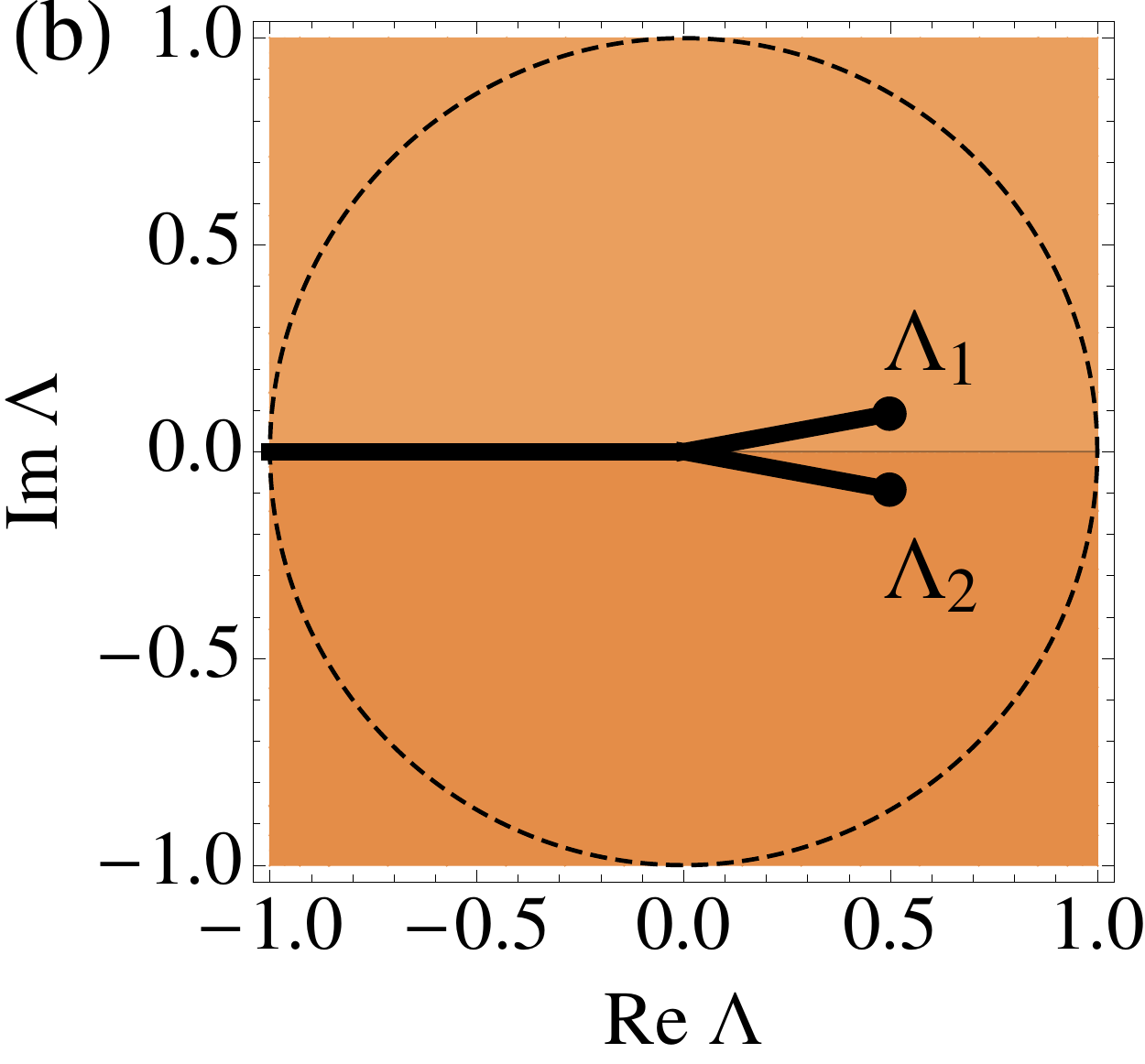}
  
  \includegraphics[width=0.23\textwidth]{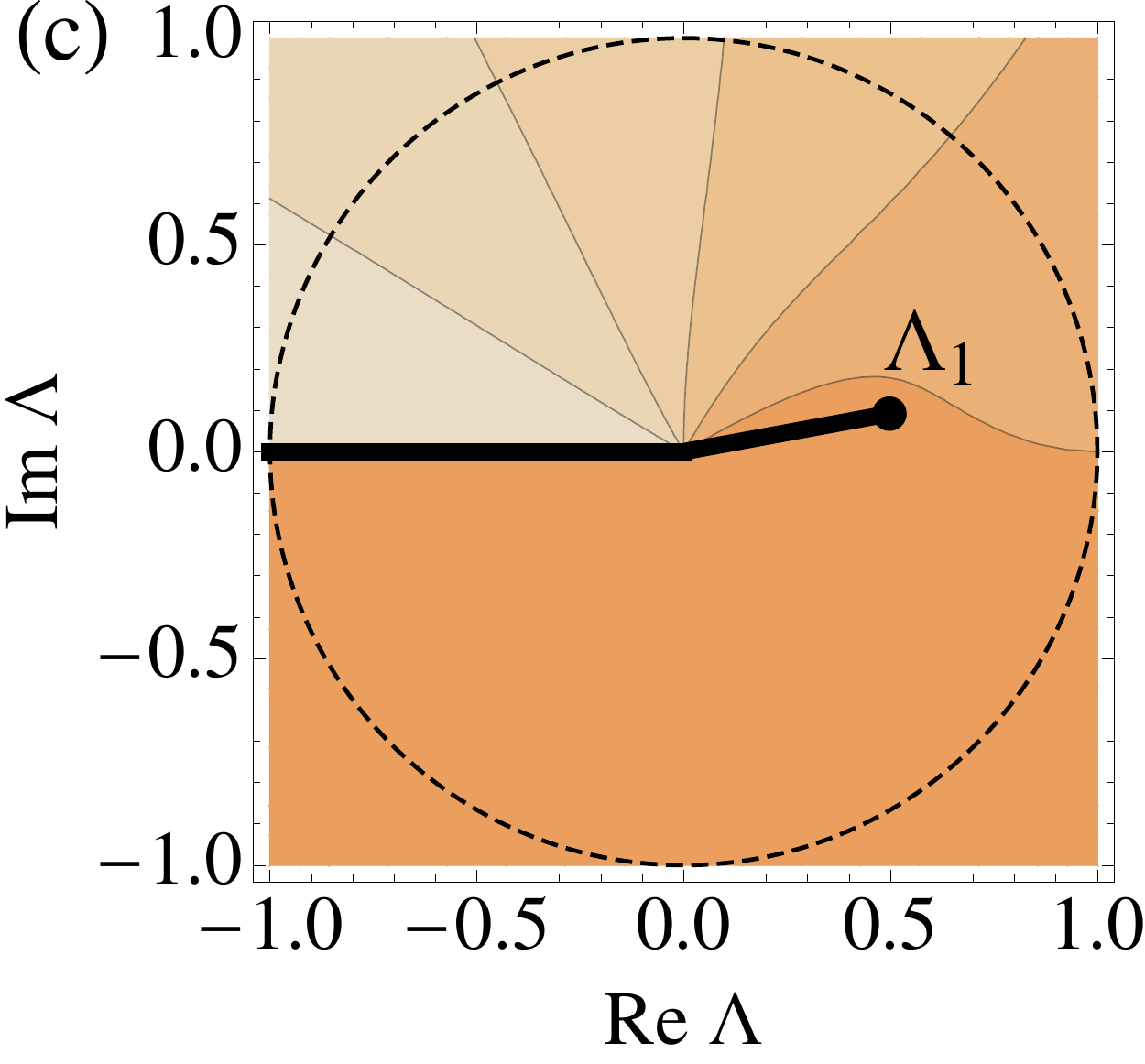}
  \hspace{0.5em}
  \hspace{0.23\textwidth}

  \caption{%
    (Color online)
    Contour plots of $\rmRe E_M$ in the $\Lambda$-plane with
    (a) $M=-1$, (b) $M=0$ and (c) $M=1$. We choose
    $\omega=\pi/6$, where a $3$EP is split into two $2$EPs indicated
    by filled 
    circles. 
    Other parameters are the same as in Fig.~\ref{fig:EP3}.
    }
  \label{fig:J_1_nu_1_12}
\end{figure}

In addition to the configuration of
these EPs, 
we need to take into account the branch cuts of quasienergies
in the $\Lambda$-plane to discuss the deformation of the adiabatic cycle.
Here we repeat the same procedure to carry out the analytic continuation
(see Sec.~\ref{sec:dEP}) to obtain the Riemann sheets of quasienergies,
as shown in Fig.~\ref{fig:J_1_nu_1_12}.
As a result, we find how the Riemann sheets of quasienergies are 
connected by the EPs and the branch cuts. 
$E_{-1}$ and $E_{0}$ degenerate at $\Lambda_2$. On the
other hand, $E_{0}$ and $E_{1}$ degenerate at $\Lambda_1$.

\begin{figure}
  \centering
  \includegraphics[width=0.33\textwidth]{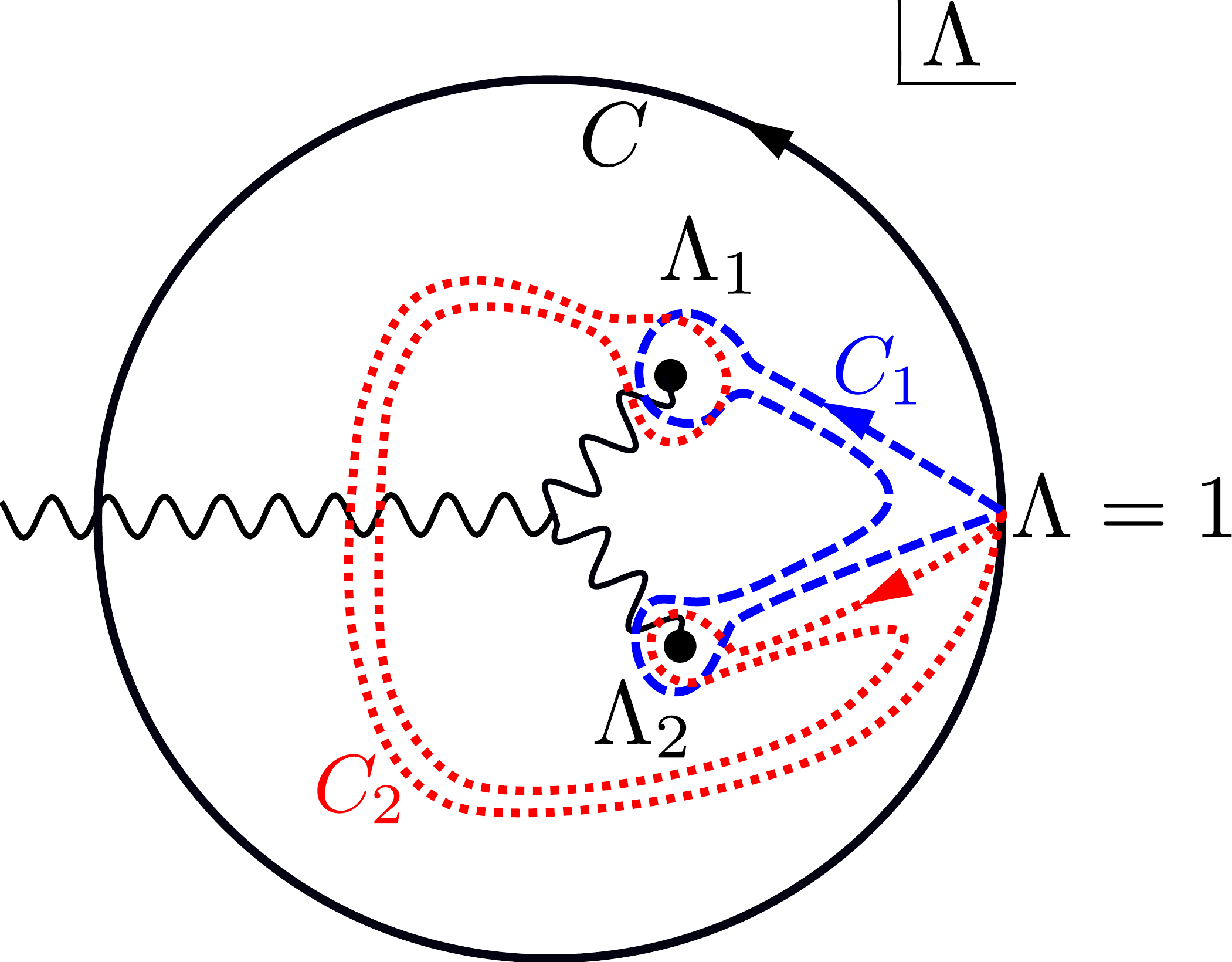}
  \caption{%
    (Color online)
    A schematic explanation of two non-Hermitian cycles 
    $C_1$ (dashed)
    and $C_2$ (dotted) that emulate a unitary cycle $C$
    in the $\Lambda$-plane
    for the quantum kicked spin $J=1$ and $\omega=\pi/6$ 
    (see, Fig.~\ref{fig:J_1_nu_1_12}).
    $C$ start from $\Lambda=1$, and encircles 
    EPs $\Lambda_1$ and $\Lambda_2$.
    }
  \label{fig:J_1emurate}
\end{figure}

There are two kinds of non-Hermitian cycles that 
emulate
the exotic 
quantum holonomy induced by the unitary cycle $C$. Typical examples
$C_1$ and  $C_2$ are shown in Fig.~\ref{fig:J_1emurate}. They are 
obtained through smooth deformations of $C$.
Along $C_1$, we first 
encircle
$\Lambda_1$, where $E_{M=0}$ and $E_{M=-1}$ are
interchanged, and then we 
encircle
$\Lambda_2$, where $E_{M=-1}$ and $E_{M=0}$
are interchanged. The composition of these permutation results in
the cyclic permutation Eq.~\eqref{eq:J1permutation}.

On the other hand, we need to take into account the presence of branch
cuts for the analysis of the cycle $C_2$. Along $C_2$, we first 
encircle
$\Lambda_2$, where $E_{M=-1}$ and $E_{M=0}$. Then, we need to come across
a branch cut. Because of this, $E_{M}$ becomes $E_{M-1}$. 
Next we 
encircle
$\Lambda_2$, and then come across the branch cut again.
Thus we conclude that the adiabatic cycle along $C_2$ 
also induces the cyclic permutation shown in Eq.~\eqref{eq:J1permutation}.

We summarize the analysis of the $J=1$ case to clarify
the correspondence between the exotic quantum holonomy and
the bifurcation of EPs. While the exotic quantum holonomy along the adiabatic cycle $C$ 
is kept intact
for $0 < \omega \le \pi$, the configuration of EPs are sensitive to $\omega$. 
There is a single $3$EP within the unit circle of the $\Lambda$-plane,
at $\omega=2\pi/3$. It is straightforward to obtain the non-Hermitian cycle, which encircle the $3$EP (e.g., $C'$ in Fig.~\ref{fig:EP3}), to emulate the exotic quantum holonomy.
As for the 
two $2$EPs case 
($\omega\ne 2\pi/3$), 
two cycles that enclose the $2$EPs are combined
to emulate the exotic quantum holonomy. 
The contributions from the 
two
$2$EPs are not 
interchangeable.
They need to be combined 
in exact order 
to correctly emulate the exotic 
quantum holonomy (see, $C_1$ and $C_2$ in Fig.~\ref{fig:J_1emurate}).

\begin{figure}[b]
  \centering

  \includegraphics[width=0.22\textwidth]{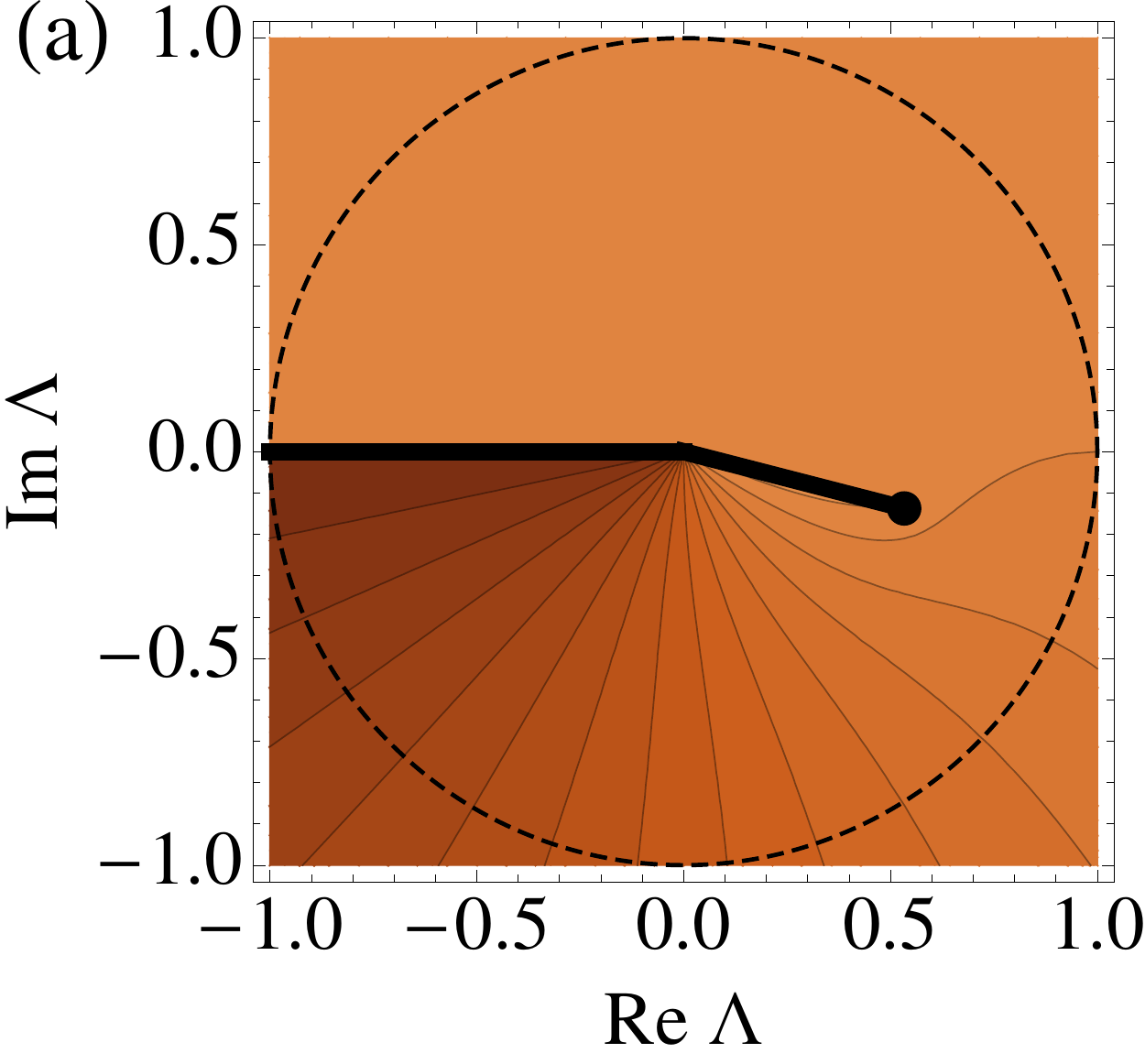}
  \hspace{0.5em}
  \includegraphics[width=0.22\textwidth]{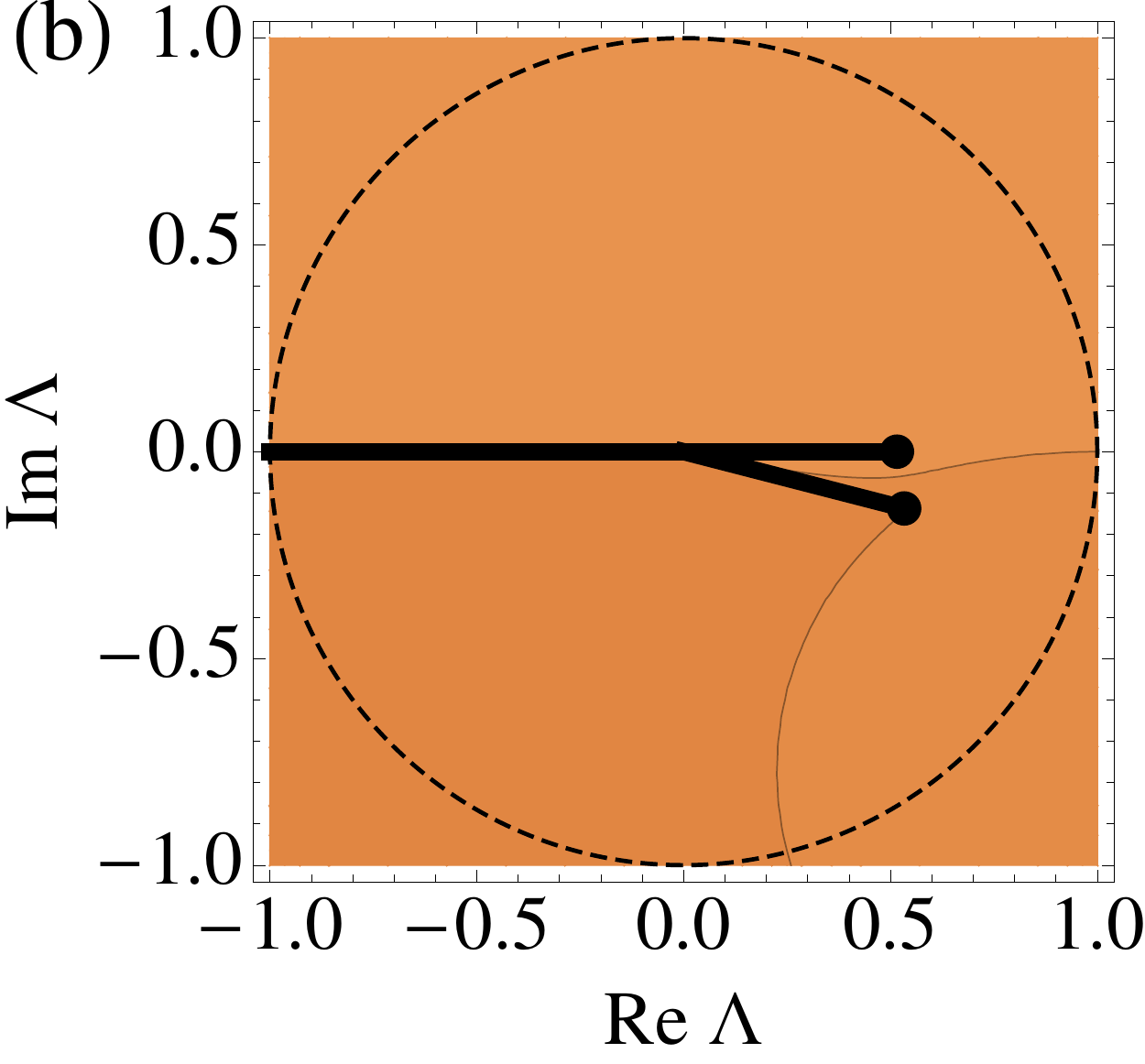}
  \\
  \includegraphics[width=0.22\textwidth]{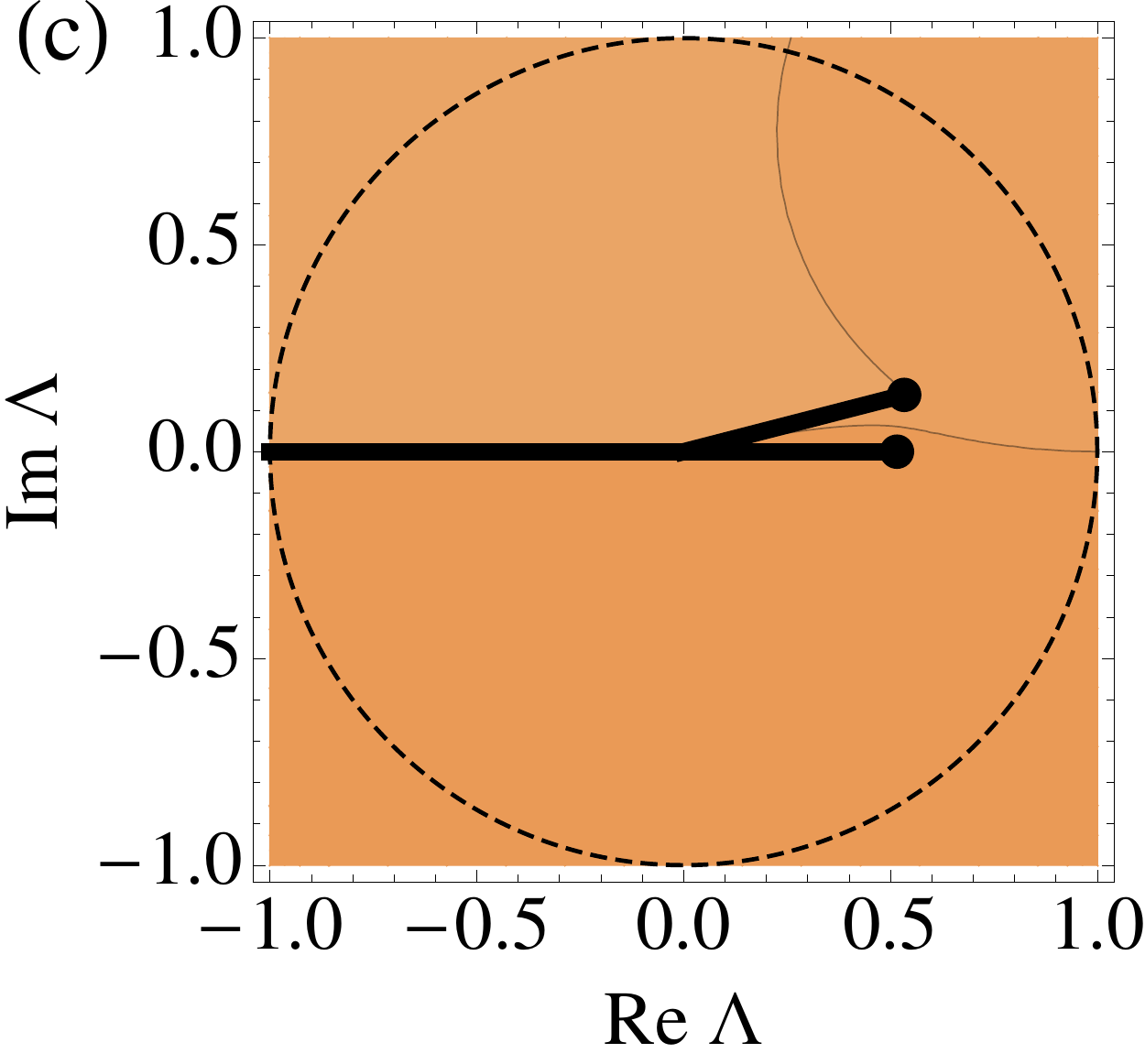}
  \hspace{0.5em}
  \includegraphics[width=0.22\textwidth]{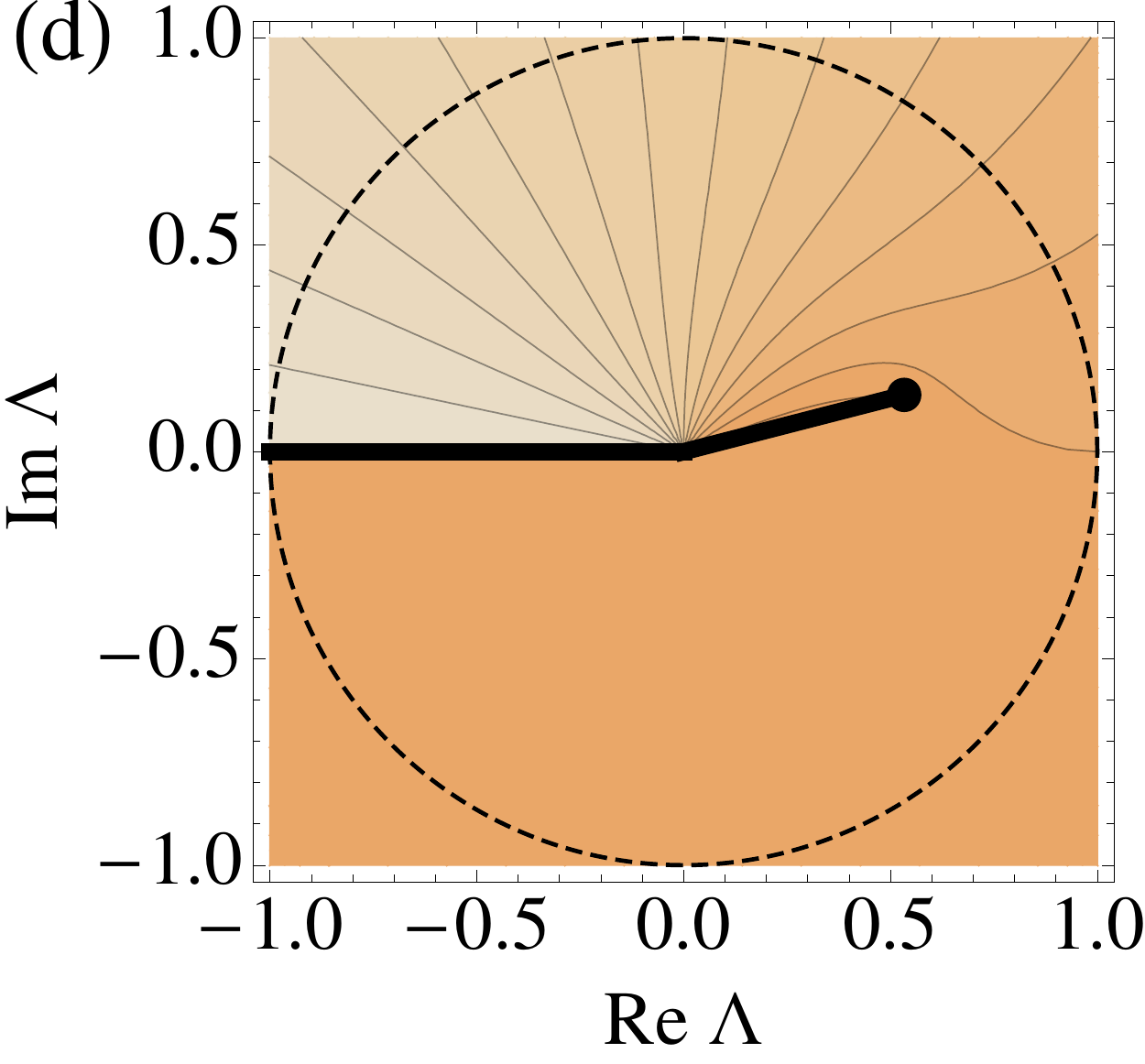}

  \caption{%
    (Color online)
    Contour plots of $\rmRe E_M$ in the $\Lambda$-plane with
    (a) $M=-3/2$, (b) $M=-1/2$, (c) $M=1/2$ and (c) $M=3/2$. 
    Other parameters are $J=3/2$ and $\omega=\pi/8$.
    There are three 
    $2$EPs, 
    which merges at $\omega=\pi/2$ (not shown here)
    to compose a $4$EP.
    Other conventions are the same as in Fig.~\ref{fig:EP3}.
    }
  \label{fig:J_3_2_nu_1_16}
\end{figure}

A similar analysis can be carried out for the quantum kicked top with 
an arbitrary $J$. We depict 
$J=3/2$ case in the vicinity of
a
$4$EP in Fig.~\ref{fig:J_3_2_nu_1_16}.

\section{Summary}
\label{sec:summary}
We 
have examined the exotic quantum holonomy in a family of quantum kicked
tops-$J$, against an adiabatic cycle $C$, where the strength of the
kick is increased.  This model exhibits the exotic quantum holonomy
for an arbitrary 
value of the parameter of the unperturbed
top $\omega$, except  at the resonant points.  
This model has a
$(2J+1)$-EP for a specific value of $\omega$. It is shown 
that an
infinitesimal perturbation can split 
$(2J+1)$-EP into $2J$ 
$2$EPs
with
$J=1$ and $3/2$.
Analytic argument is provided for $J=1$.
We 
have observed an intricate interplay of the exotic quantum holonomy and
EPs both the $(2J+1)$-EP case and the $2$EPs case.

\section*{Acknowledgments}
This research was supported by NRF funded by the Ministry of Science, ICT 
and Future Planning (2013R1A1A2011438),
and by the Japan Ministry of Education, Culture, Sports, Science and Technology under the Grant number 24540412.


\begin{thebibliography}{36}
\expandafter\ifx\csname natexlab\endcsname\relax\def\natexlab#1{#1}\fi
\expandafter\ifx\csname bibnamefont\endcsname\relax
  \def\bibnamefont#1{#1}\fi
\expandafter\ifx\csname bibfnamefont\endcsname\relax
  \def\bibfnamefont#1{#1}\fi
\expandafter\ifx\csname citenamefont\endcsname\relax
  \def\citenamefont#1{#1}\fi
\expandafter\ifx\csname url\endcsname\relax
  \def\url#1{\texttt{#1}}\fi
\expandafter\ifx\csname urlprefix\endcsname\relax\def\urlprefix{URL }\fi
\providecommand{\bibinfo}[2]{#2}
\providecommand{\eprint}[2][]{\url{#2}}

\bibitem[{\citenamefont{Kato}(1950)}]{Kato-JPSJ-5-435}
\bibinfo{author}{\bibfnamefont{T.}~\bibnamefont{Kato}}, \bibinfo{journal}{J.
  Phys. Soc. Japan} \textbf{\bibinfo{volume}{5}}, \bibinfo{pages}{435}
  (\bibinfo{year}{1950}).

\bibitem[{\citenamefont{Berry}(1984)}]{Berry-PRSLA-392-45}
\bibinfo{author}{\bibfnamefont{M.~V.} \bibnamefont{Berry}},
  \bibinfo{journal}{Proc. Roy. Soc. London} \textbf{\bibinfo{volume}{A 392}},
  \bibinfo{pages}{45} (\bibinfo{year}{1984}).

\bibitem[{\citenamefont{Wilczek and Zee}(1984)}]{Wilczek-PRL-52-2111}
\bibinfo{author}{\bibfnamefont{F.}~\bibnamefont{Wilczek}} \bibnamefont{and}
  \bibinfo{author}{\bibfnamefont{A.}~\bibnamefont{Zee}},
  \bibinfo{journal}{Phys. Rev. Lett.} \textbf{\bibinfo{volume}{52}},
  \bibinfo{pages}{2111} (\bibinfo{year}{1984}).

\bibitem[{\citenamefont{Cheon}(1998)}]{Cheon-PLA-248-285}
\bibinfo{author}{\bibfnamefont{T.}~\bibnamefont{Cheon}},
  \bibinfo{journal}{Phys. Lett. A} \textbf{\bibinfo{volume}{248}},
  \bibinfo{pages}{285} (\bibinfo{year}{1998}).

\bibitem[{\citenamefont{Tanaka and Miyamoto}(2007)}]{Tanaka-PRL-98-160407}
\bibinfo{author}{\bibfnamefont{A.}~\bibnamefont{Tanaka}} \bibnamefont{and}
  \bibinfo{author}{\bibfnamefont{M.}~\bibnamefont{Miyamoto}},
  \bibinfo{journal}{Phys. Rev. Lett.} \textbf{\bibinfo{volume}{98}},
  \bibinfo{pages}{160407} (\bibinfo{year}{2007}).

\bibitem[{\citenamefont{Miyamoto and Tanaka}(2007)}]{Miyamoto-PRA-76-042115}
\bibinfo{author}{\bibfnamefont{M.}~\bibnamefont{Miyamoto}} \bibnamefont{and}
  \bibinfo{author}{\bibfnamefont{A.}~\bibnamefont{Tanaka}},
  \bibinfo{journal}{Phys. Rev. A} \textbf{\bibinfo{volume}{76}},
  \bibinfo{pages}{042115} (\bibinfo{year}{2007}).

\bibitem[{\citenamefont{Cheon and Tanaka}(2009)}]{Cheon-EPL-85-20001}
\bibinfo{author}{\bibfnamefont{T.}~\bibnamefont{Cheon}} \bibnamefont{and}
  \bibinfo{author}{\bibfnamefont{A.}~\bibnamefont{Tanaka}},
  \bibinfo{journal}{Europhys. Lett.} \textbf{\bibinfo{volume}{85}},
  \bibinfo{pages}{20001} (\bibinfo{year}{2009}).

\bibitem[{\citenamefont{Cheon et~al.}(2009)\citenamefont{Cheon, Tanaka, and
  Kim}}]{Cheon-PLA-374-144}
\bibinfo{author}{\bibfnamefont{T.}~\bibnamefont{Cheon}},
  \bibinfo{author}{\bibfnamefont{A.}~\bibnamefont{Tanaka}}, \bibnamefont{and}
  \bibinfo{author}{\bibfnamefont{S.~W.} \bibnamefont{Kim}},
  \bibinfo{journal}{Phys. Lett. A} \textbf{\bibinfo{volume}{374}},
  \bibinfo{pages}{144} (\bibinfo{year}{2009}).

\bibitem[{\citenamefont{Tanaka and Cheon}(2010)}]{Tanaka-PRA-82-022104}
\bibinfo{author}{\bibfnamefont{A.}~\bibnamefont{Tanaka}} \bibnamefont{and}
  \bibinfo{author}{\bibfnamefont{T.}~\bibnamefont{Cheon}},
  \bibinfo{journal}{Phys. Rev. A} \textbf{\bibinfo{volume}{82}},
  \bibinfo{pages}{022104} (\bibinfo{year}{2010}).

\bibitem[{\citenamefont{Tanaka et~al.}(2011)\citenamefont{Tanaka, Kim, and
  Cheon}}]{Tanaka-EPL-96-10005}
\bibinfo{author}{\bibfnamefont{A.}~\bibnamefont{Tanaka}},
  \bibinfo{author}{\bibfnamefont{S.~W.} \bibnamefont{Kim}}, \bibnamefont{and}
  \bibinfo{author}{\bibfnamefont{T.}~\bibnamefont{Cheon}},
  \bibinfo{journal}{Europhys. Lett.} \textbf{\bibinfo{volume}{96}},
  \bibinfo{pages}{10005} (\bibinfo{year}{2011}).

\bibitem[{\citenamefont{Yonezawa et~al.}(2013)\citenamefont{Yonezawa, Tanaka,
  and Cheon}}]{Yonezawa-PRA-87-062113}
\bibinfo{author}{\bibfnamefont{N.}~\bibnamefont{Yonezawa}},
  \bibinfo{author}{\bibfnamefont{A.}~\bibnamefont{Tanaka}}, \bibnamefont{and}
  \bibinfo{author}{\bibfnamefont{T.}~\bibnamefont{Cheon}},
  \bibinfo{journal}{Phys. Rev. A} \textbf{\bibinfo{volume}{87}},
  \bibinfo{pages}{062113} (\bibinfo{year}{2013}).

\bibitem[{\citenamefont{Farhi et~al.}(2000)\citenamefont{Farhi, Goldstone,
  Gutmann, and Sipser}}]{Farhi-quant-ph-0001106}
\bibinfo{author}{\bibfnamefont{E.}~\bibnamefont{Farhi}},
  \bibinfo{author}{\bibfnamefont{J.}~\bibnamefont{Goldstone}},
  \bibinfo{author}{\bibfnamefont{S.}~\bibnamefont{Gutmann}}, \bibnamefont{and}
  \bibinfo{author}{\bibfnamefont{M.}~\bibnamefont{Sipser}}
  (\bibinfo{year}{2000}).

\bibitem[{\citenamefont{Tanaka and Nemoto}(2010)}]{Tanaka-PRA-81-022320}
\bibinfo{author}{\bibfnamefont{A.}~\bibnamefont{Tanaka}} \bibnamefont{and}
  \bibinfo{author}{\bibfnamefont{K.}~\bibnamefont{Nemoto}},
  \bibinfo{journal}{Phys. Rev. A} \textbf{\bibinfo{volume}{81}},
  \bibinfo{pages}{022320} (\bibinfo{year}{2010}).

\bibitem[{\citenamefont{Berry et~al.}(1979)\citenamefont{Berry, Balazs, Tabor,
  and Voros}}]{Berry-AP-122-26}
\bibinfo{author}{\bibfnamefont{M.}~\bibnamefont{Berry}},
  \bibinfo{author}{\bibfnamefont{N.}~\bibnamefont{Balazs}},
  \bibinfo{author}{\bibfnamefont{M.}~\bibnamefont{Tabor}}, \bibnamefont{and}
  \bibinfo{author}{\bibfnamefont{A.}~\bibnamefont{Voros}},
  \bibinfo{journal}{Ann. Phys. (NY)} \textbf{\bibinfo{volume}{122}},
  \bibinfo{pages}{26} (\bibinfo{year}{1979}).

\bibitem[{\citenamefont{Zel{'}dovich}(1967)}]{Zeldovich-JETP-24-1006}
\bibinfo{author}{\bibfnamefont{Y.~B.} \bibnamefont{Zel{'}dovich}},
  \bibinfo{journal}{Sov. Phys.--JETP} \textbf{\bibinfo{volume}{24}},
  \bibinfo{pages}{1006} (\bibinfo{year}{1967}).

\bibitem[{\citenamefont{Sadgrove et~al.}(2007)\citenamefont{Sadgrove,
  Horikoshi, Sekimura, and Nakagawa}}]{Sadgrove-PRL-99-043002}
\bibinfo{author}{\bibfnamefont{M.}~\bibnamefont{Sadgrove}},
  \bibinfo{author}{\bibfnamefont{M.}~\bibnamefont{Horikoshi}},
  \bibinfo{author}{\bibfnamefont{T.}~\bibnamefont{Sekimura}}, \bibnamefont{and}
  \bibinfo{author}{\bibfnamefont{K.}~\bibnamefont{Nakagawa}},
  \bibinfo{journal}{Phys. Rev. Lett.} \textbf{\bibinfo{volume}{99}},
  \bibinfo{pages}{043002} (\bibinfo{year}{2007}).

\bibitem[{\citenamefont{Chab{\'e} et~al.}(2008)\citenamefont{Chab{\'e},
  Lemari{\'e}, Gr{\'e}maud, Delande, Szriftgiser, and
  Garreau}}]{Chabe-PRL-101-255702}
\bibinfo{author}{\bibfnamefont{J.}~\bibnamefont{Chab{\'e}}},
  \bibinfo{author}{\bibfnamefont{G.}~\bibnamefont{Lemari{\'e}}},
  \bibinfo{author}{\bibfnamefont{B.}~\bibnamefont{Gr{\'e}maud}},
  \bibinfo{author}{\bibfnamefont{D.}~\bibnamefont{Delande}},
  \bibinfo{author}{\bibfnamefont{P.}~\bibnamefont{Szriftgiser}},
  \bibnamefont{and} \bibinfo{author}{\bibfnamefont{J.~C.}
  \bibnamefont{Garreau}}, \bibinfo{journal}{Phys. Rev. Lett.}
  \textbf{\bibinfo{volume}{101}}, \bibinfo{pages}{255702}
  (\bibinfo{year}{2008}).

\bibitem[{\citenamefont{Chaudhury et~al.}(2009)\citenamefont{Chaudhury, Smith,
  Anderson, Chose, and Jessen}}]{Chaudhry-Nature-461-768}
\bibinfo{author}{\bibfnamefont{S.}~\bibnamefont{Chaudhury}},
  \bibinfo{author}{\bibfnamefont{A.}~\bibnamefont{Smith}},
  \bibinfo{author}{\bibfnamefont{B.~E.} \bibnamefont{Anderson}},
  \bibinfo{author}{\bibfnamefont{S.}~\bibnamefont{Chose}}, \bibnamefont{and}
  \bibinfo{author}{\bibfnamefont{P.}~\bibnamefont{Jessen}},
  \bibinfo{journal}{Nature} \textbf{\bibinfo{volume}{461}},
  \bibinfo{pages}{768} (\bibinfo{year}{2009}).

\bibitem[{\citenamefont{Morello et~al.}(2010)\citenamefont{Morello, Pla,
  Zwanenburg, Chan, Tan, Huebl, Möttönen, Nugroho, Yang, van Donkelaar
  et~al.}}]{Morello-Nature-467-687}
\bibinfo{author}{\bibfnamefont{A.}~\bibnamefont{Morello}} 
  \bibnamefont{et~al.}, \bibinfo{journal}{Nature}
  \textbf{\bibinfo{volume}{467}}, \bibinfo{pages}{687} (\bibinfo{year}{2010}).

\bibitem[{\citenamefont{Kato}(1980)}]{KatoExceptionalPoint}
\bibinfo{author}{\bibfnamefont{T.}~\bibnamefont{Kato}},
  \emph{\bibinfo{title}{Perturbation Theory for Linear Operators}}
  (\bibinfo{publisher}{Springer-Verlag}, \bibinfo{address}{Berlin},
  \bibinfo{year}{1980}), chap.~\bibinfo{chapter}{II},
  \bibinfo{edition}{corrected printing of the second} ed.

\bibitem[{\citenamefont{Uzdin et~al.}(2011)\citenamefont{Uzdin, Mailybaev, and
  Moiseyev}}]{Uzdin-JPA-44-435302}
\bibinfo{author}{\bibfnamefont{R.}~\bibnamefont{Uzdin}},
  \bibinfo{author}{\bibfnamefont{A.}~\bibnamefont{Mailybaev}},
  \bibnamefont{and} \bibinfo{author}{\bibfnamefont{N.}~\bibnamefont{Moiseyev}},
  \bibinfo{journal}{Journal of Physics A: Mathematical and Theoretical}
  \textbf{\bibinfo{volume}{44}}, \bibinfo{pages}{435302}
  (\bibinfo{year}{2011}).

\bibitem[{\citenamefont{Berry and Uzdin}(2011)}]{Berry-JPA-44-435303}
\bibinfo{author}{\bibfnamefont{M.~V.} \bibnamefont{Berry}} \bibnamefont{and}
  \bibinfo{author}{\bibfnamefont{R.}~\bibnamefont{Uzdin}},
  \bibinfo{journal}{Journal of Physics A: Mathematical and Theoretical}
  \textbf{\bibinfo{volume}{44}}, \bibinfo{pages}{435303}
  (\bibinfo{year}{2011}).

\bibitem[{\citenamefont{Uzdin and Moiseyev}(2012)}]{Uzdin-PRA-85-031804}
\bibinfo{author}{\bibfnamefont{R.}~\bibnamefont{Uzdin}} \bibnamefont{and}
  \bibinfo{author}{\bibfnamefont{N.}~\bibnamefont{Moiseyev}},
  \bibinfo{journal}{Physical Review A} \textbf{\bibinfo{volume}{85}},
  \bibinfo{pages}{031804} (\bibinfo{year}{2012}).

\bibitem[{\citenamefont{Kim et~al.}(2010)\citenamefont{Kim, Cheon, and
  Tanaka}}]{Kim-PLA-374-1958}
\bibinfo{author}{\bibfnamefont{S.~W.} \bibnamefont{Kim}},
  \bibinfo{author}{\bibfnamefont{T.}~\bibnamefont{Cheon}}, \bibnamefont{and}
  \bibinfo{author}{\bibfnamefont{A.}~\bibnamefont{Tanaka}},
  \bibinfo{journal}{Phys. Lett. A} \textbf{\bibinfo{volume}{374}},
  \bibinfo{pages}{1958} (\bibinfo{year}{2010}).

\bibitem[{\citenamefont{Ryu et~al.}(2012)\citenamefont{Ryu, Lee, and
  Kim}}]{Ryu-PRA-85-042101}
\bibinfo{author}{\bibfnamefont{J.-W.} \bibnamefont{Ryu}},
  \bibinfo{author}{\bibfnamefont{S.-Y.} \bibnamefont{Lee}}, \bibnamefont{and}
  \bibinfo{author}{\bibfnamefont{S.~W.} \bibnamefont{Kim}},
  \bibinfo{journal}{Phys. Rev. A} \textbf{\bibinfo{volume}{85}},
  \bibinfo{pages}{042101} (\bibinfo{year}{2012}).

\bibitem[{\citenamefont{Lee et~al.}(2012)\citenamefont{Lee, Ryu, Kim, and
  Chung}}]{Lee-PRA-85-064103}
\bibinfo{author}{\bibfnamefont{S.-Y.} \bibnamefont{Lee}},
  \bibinfo{author}{\bibfnamefont{J.-W.} \bibnamefont{Ryu}},
  \bibinfo{author}{\bibfnamefont{S.~W.} \bibnamefont{Kim}}, \bibnamefont{and}
  \bibinfo{author}{\bibfnamefont{Y.}~\bibnamefont{Chung}},
  \bibinfo{journal}{Phys. Rev. A} \textbf{\bibinfo{volume}{85}},
  \bibinfo{pages}{064103} (\bibinfo{year}{2012}).

\bibitem[{\citenamefont{Tanaka et~al.}(2013)\citenamefont{Tanaka, Yonezawa, and
  Cheon}}]{Tanaka-JPA-46-315302}
\bibinfo{author}{\bibfnamefont{A.}~\bibnamefont{Tanaka}},
  \bibinfo{author}{\bibfnamefont{N.}~\bibnamefont{Yonezawa}}, \bibnamefont{and}
  \bibinfo{author}{\bibfnamefont{T.}~\bibnamefont{Cheon}},
  \bibinfo{journal}{Journal of Physics A: Mathematical and Theoretical}
  \textbf{\bibinfo{volume}{46}}, \bibinfo{pages}{315302}
  (\bibinfo{year}{2013}).

\bibitem[{\citenamefont{Heiss}(2008)}]{Heiss-JPA-41-244010}
\bibinfo{author}{\bibfnamefont{W.~D.} \bibnamefont{Heiss}},
  \bibinfo{journal}{J. Phys. A.} \textbf{\bibinfo{volume}{41}},
  \bibinfo{pages}{244010} (\bibinfo{year}{2008}).

\bibitem[{\citenamefont{Graefe et~al.}(2008)\citenamefont{Graefe, G{\"u}nther,
  Korsch, and Niederle}}]{Graefe-JPA-41-255206}
\bibinfo{author}{\bibfnamefont{E.~M.} \bibnamefont{Graefe}},
  \bibinfo{author}{\bibfnamefont{U.}~\bibnamefont{G{\"u}nther}},
  \bibinfo{author}{\bibfnamefont{H.~J.} \bibnamefont{Korsch}},
  \bibnamefont{and} \bibinfo{author}{\bibfnamefont{A.~E.}
  \bibnamefont{Niederle}}, \bibinfo{journal}{J. Phys. A.}
  \textbf{\bibinfo{volume}{41}}, \bibinfo{pages}{255206}
  (\bibinfo{year}{2008}).

\bibitem[{\citenamefont{Combescure}(1990)}]{Combesqure-JSP-59-679}
\bibinfo{author}{\bibfnamefont{M.}~\bibnamefont{Combescure}},
  \bibinfo{journal}{J. Stat. Phys.} \textbf{\bibinfo{volume}{59}},
  \bibinfo{pages}{679} (\bibinfo{year}{1990}).

\bibitem[{\citenamefont{Ghose et~al.}(2008)\citenamefont{Ghose, Stock, Jessen,
  Lal, and Silberfarb}}]{Chose-PRA-78-042318}
\bibinfo{author}{\bibfnamefont{S.}~\bibnamefont{Ghose}},
  \bibinfo{author}{\bibfnamefont{R.}~\bibnamefont{Stock}},
  \bibinfo{author}{\bibfnamefont{P.}~\bibnamefont{Jessen}},
  \bibinfo{author}{\bibfnamefont{R.}~\bibnamefont{Lal}}, \bibnamefont{and}
  \bibinfo{author}{\bibfnamefont{A.}~\bibnamefont{Silberfarb}},
  \bibinfo{journal}{Phys. Rev. A} \textbf{\bibinfo{volume}{78}},
  \bibinfo{pages}{042318} (\bibinfo{year}{2008}).

\bibitem[{\citenamefont{Srinivas and Davies}(1981)}]{Srinivas-OptAc-28-981}
\bibinfo{author}{\bibfnamefont{M.}~\bibnamefont{Srinivas}} \bibnamefont{and}
  \bibinfo{author}{\bibfnamefont{E.}~\bibnamefont{Davies}},
  \bibinfo{journal}{Optica Acta} \textbf{\bibinfo{volume}{28}},
  \bibinfo{pages}{981} (\bibinfo{year}{1981}).

\bibitem[{\citenamefont{Mohseni et~al.}(2008)\citenamefont{Mohseni, Rebentrost,
  Lloyd, and Aspuru-Guzik}}]{Mohseni-JCP-129-174106}
\bibinfo{author}{\bibfnamefont{M.}~\bibnamefont{Mohseni}},
  \bibinfo{author}{\bibfnamefont{P.}~\bibnamefont{Rebentrost}},
  \bibinfo{author}{\bibfnamefont{S.}~\bibnamefont{Lloyd}}, \bibnamefont{and}
  \bibinfo{author}{\bibfnamefont{A.}~\bibnamefont{Aspuru-Guzik}},
  \bibinfo{journal}{J. Chem. Phys.} \textbf{\bibinfo{volume}{129}},
  \bibinfo{pages}{174106} (\bibinfo{year}{2008}).

\bibitem[{bio()}]{biorthogonal}
\bibinfo{note}{See, e.g., N. Moiseyev, {\it Non-Hermitian Quantum Mechanics}
  (Cambridge Univ. Press, New York, 2011)}.

\bibitem[{\citenamefont{Olver et~al.}(2010)\citenamefont{Olver, Lozier,
  Boisvert, and Clark}}]{Cardano}
\bibinfo{editor}{\bibfnamefont{F.~W.~J.} \bibnamefont{Olver}},
  \bibinfo{editor}{\bibfnamefont{D.~W.} \bibnamefont{Lozier}},
  \bibinfo{editor}{\bibfnamefont{R.~F.} \bibnamefont{Boisvert}},
  \bibnamefont{and} \bibinfo{editor}{\bibfnamefont{C.~W.} \bibnamefont{Clark}},
  eds., \emph{\bibinfo{title}{NIST Handbook of Mathematical Functions}}
  (\bibinfo{publisher}{Cambridge University Press},
  \bibinfo{address}{Cambridge}, \bibinfo{year}{2010}),
  chap.~\bibinfo{chapter}{1}.

\bibitem[{\citenamefont{van~der Waerdern}(1949)}]{Resultant}
\bibinfo{author}{\bibfnamefont{B.}~\bibnamefont{van~der Waerdern}},
  \emph{\bibinfo{title}{Moderne Algebra}} (\bibinfo{publisher}{Frederick Ungar
  Publishing}, \bibinfo{address}{New York}, \bibinfo{year}{1949}),
  vol.~\bibinfo{volume}{I}, chap.~\bibinfo{chapter}{4}.

\end{thebibliography}



\end{document}